\documentclass[11pt,a4paper]{article}
\usepackage{amssymb}
\usepackage{amsthm}

\usepackage{amsmath}
\usepackage{tikz}
\usepackage{algcompatible}
\usepackage{algorithm}
\usepackage[justification=centering]{caption}
\usetikzlibrary{arrows}
\usetikzlibrary{automata}
\usetikzlibrary{positioning} 
\usetikzlibrary{decorations.pathmorphing}

\newtheorem{lemma1}{Lemma}[section]
\newtheorem{theorem1}[lemma1]{Theorem}
\newtheorem{proposition1}[lemma1]{Proposition}

\newcommand{\scp}[1]{{\sc #1}}
\newcommand{\sfp}[1]{{\tt #1}}

\begin{document}
\title{\bf Profile-based optimal matchings in the Student/Project Allocation problem\thanks{A preliminary version of this paper appeared in the proceedings of IWOCA 2014: the 25th International Workshop on Combinatorial Algorithms.}}
\author{Augustine Kwanashie$^1$, Robert W. Irving$^1$,\\David F. Manlove$^{1,}$\thanks{Supported by Engineering and Physical Sciences Research Council grant EP/K010042/1.  Corresponding author.  Email {\tt david.manlove@glasgow.ac.uk}.}~ and Colin T.S. Sng$^{2,}$\thanks{Work done while at the School of Computing Science, University of Glasgow.}
\\
\\
\small $^1$ School of Computing Science, University of Glasgow, UK
%\\ Email: a.kwanashie.1@research.gla.ac.uk, rob.irving@glasgow.ac.uk david.manlove@glasgow.ac.uk.\\
\\
\small $^2$ Amazon.com, Inc., Texas, USA  %Email colinsng@amazon.com.
}
\date{}
\maketitle
\begin{abstract}
In the \emph{Student / Project Allocation problem {\sc(spa)}} we seek to assign students to individual or group projects offered by lecturers. Students provide a list of projects they find acceptable in order of preference. Each student can be assigned to at most one project and there are constraints on the maximum number of students that can be assigned to each project and lecturer. We seek matchings of students to projects that are optimal with respect to \emph{profile}, which is a vector whose $r$th component indicates how many students have their $r$th-choice project. We present an efficient algorithm for finding a \emph{greedy maximum matching} in the {\sc spa} context -- this is a maximum matching whose profile is lexicographically maximum. We then show how to adapt this algorithm to find a \emph{generous maximum matching} -- this is a matching whose reverse profile is lexicographically minimum. Our algorithms involve finding optimal flows in networks. 
We demonstrate how this approach can allow for additional constraints, such as lecturer lower quotas, to be handled flexibly. Finally we present results obtained from an empirical evaluation of the algorithms. 
\end{abstract}

%=======================================================================================================================
%=======================================================================================================================
%\input{students_project.tex}

\label{ch_spa1}
\section{Introduction}
In most academic programmes students are usually required to take up individual or group projects offered by lecturers. Students may be required to rank a subset of the projects they find acceptable in order of preference. Each project is offered by a unique lecturer who may also be allowed to rank the projects she offers or the students who are interested in taking her projects in order of preference.  Each student can be assigned to at most one project and there are usually constraints on the maximum number of students that can be assigned to each project and lecturer. 
The problem then is to assign students to projects in a manner that satisfies these capacity constraints while taking into account the preferences of the students and lecturers involved. This problem has been described in the literature as the \emph{Student-Project Allocation problem {\sc (spa)}} \cite{AIM07,MO08,AM09,IMY12}.  
Variants of \scp{spa} also exist in which \emph{lower quotas} are assigned to projects and/or lecturers. These lower quotas indicate  the minimum number of students to be assigned to each project and lecturer.

Although described in an academic context, applications of {\sc spa} need not be limited to assigning students to projects but may extend to other scenarios, such as the assignment of employees to posts in a company where available posts are offered by various departments.  Applications of {\sc spa} in an academic context can be found at the University of Glasgow \cite{MU14}, the University of York \cite{Dye01,Kaz02,Tho03}, the University of Southampton \cite{AB03,HSVS05} and the Geneva School of Business Administration \cite{VSc13}. 
As previously stated, it is widely accepted that matching problems (like {\sc spa}) are best solved by centralised matching schemes where agents submit their preferences and a central authority computes an optimal matching that satisfies all the specified criteria \cite{GI89}. Moreover the potentially large number of students and projects involved in these schemes motivates the need to discover efficient algorithms for finding optimal matchings.

%In {\sc spa}, students are always required to provide preference lists containing projects. However, variants of the problem may be defined depending on the presence and nature of lecturer preference lists. Some variants of {\sc spa}, as discussed in Section \ref{two-sided}, require both students and lecturers to provide strictly ordered preference lists. In these cases, the well-known stability criterion is applied on the matchings produced. In other variants of \scp{spa}, which we introduce in Section \ref{one-sided}, only students are required to produce preference lists in which case other feasibility and optimality criteria need to be considered. In general, \scp{spa} models may vary from one application of the problem to another. In Section \ref{other-models} we discuss some of these variants that appear in the literature.

%--------------------------------------------------------------------------------------------------------------------------------------------------
\subsection{Two-sided preferences and stability} 
\label{two-sided}
In {\sc spa}, students are always required to provide preference lists over projects. However, variants of the problem may be defined depending on the presence and nature of lecturer preference lists. Some variants of {\sc spa} require both students and lecturers to provide preference lists. These variants include: (i) the \emph{Student/Project Allocation problem with lecturer preferences over Students {\sc (spa-s)}} \cite{AIM07} which requires each lecturer to rank the students who find at least one of her offered projects acceptable, in  order of preference, (ii) the \emph{Student/Project Allocation problem with lecturer preferences over Projects {\sc (spa-p)}} \cite{MO08,IMY12} which involves lecturers ranking the projects they offer in order of preference and (iii) the \emph{Student/Project Allocation problem with lecturer preferences over Student-Project pairs {\sc (spa-(s,p))}} \cite{AIM07,AM09} where lecturers rank student-project pairs in order of preference.
These variants of {\sc spa} have been studied in the context of  the well-known \emph{stability} solution criterion for matching problems \cite{GI89}. The general stability objective is to produce a matching $M$ in which no student-project pair that are not currently matched in $M$ can simultaneously improve by being paired together (thus in the process potentially abandoning their partners in $M$). A full description of the results relating to these {\sc spa} variants can be found in \cite{Man13}.

%--------------------------------------------------------------------------------------------------------------------------------------------------
\subsection{One-sided preferences and profile-based optimality}
\label{one-sided}
In many practical {\sc spa} applications it is considered appropriate to allow only students to submit preferences over projects. When preferences are specified by only one set of agents in a two-sided matching problem, the notion of stability becomes irrelevant.  This motivates the need to adopt alternative solution criteria when lecturer preferences are not allowed. In this subsection we describe some of these solution criteria and briefly present results relating to them. These criteria consider the size of the matchings produced as well as the satisfaction of the students involved.

When preference lists of lecturers are absent, the {\sc spa} problem becomes a two-sided matching problem with one-sided preferences. We assume students' preference lists can contain ties in these \scp{spa} variants.  Various optimality criteria for such problems have been studied in the literature \cite{Man13}. Some of these criteria depend on the \emph{profile} or the \emph{cost} of a matching. In the {\sc spa} context, the \emph{profile} of a matching is a vector whose $r$th component indicates the number of students obtaining their $r$th-choice project in the matching. The \emph{cost} of a matching (w.r.t. the students) is the sum of the ranks of the assigned projects in the students' preference lists (that is, the sum of $rx_r$ taken over all components $r$ of the profile, where $x_r$ is the $r$th component value).

A \emph{minimum cost maximum matching} is a maximum cardinality matching with minimum cost.   A \emph{rank-maximal matching} is a matching that has lexicographically maximum profile \cite{IKMMP06,Irv03}. That is the maximum number of students are assigned to their first-choice project and subject to this, the maximum number of students are assigned to their second choice project and so on. However a rank maximal matching need not be a maximum matching in the given instance (see, e.g., \cite[p.43]{Man13}). 
Since it is usually important to match as many students as possible, we may first optimise the size of the matching before considering student satisfaction. Thus we define a \emph{greedy maximum matching} \cite{Irv06,MM06,HK13} to be a maximum matching that has lexicographically maximum profile. The intuition behind both rank-maximal and greedy maximum matchings is to maximize the number of students matched with higher ranked projects. This may lead to some students being matched to projects that are relatively low on their preference lists. An alternative approach is to find a \emph{generous maximum matching} which is a maximum matching in which the minimum number of students are matched to their $R$th-choice project (where $R$ is the maximum length of any students' preference list) and subject to this, the minimum number of students are matched to their $(R-1)$th-choice project and so on. Greedy and generous maximum matchings have been used to assign students to projects in the School of Computing Science, and students to elective courses in the School of Medicine, both at the University of Glasgow, since 2007. Figure \ref{spa_instance} shows a  sample {\sc spa} instance with greedy and generous maximum matchings, namely $M_1 = \{(s_1, p_3), (s_2, p_1), (s_3, p_2)\}$ and $M_2 = \{(s_1, p_2), (s_2, p_1), (s_3, p_3)\}$ respectively.

\begin{figure}[t]
\centering
\small
\begin{minipage}[b]{0.6\linewidth}
\begin{align*}
&\mbox{students' preferences:} ~~~~~~~~~&&\mbox{lecturers' offerings:}\\
&s_1: p_1~~~p_2~~~p_3  &&l_1: \{p_1, p_2\} \\
&s_2: p_1 &&l_2: \{p_3\} \\
&s_3: p_2~~~p_3 &&\mbox{project capacities: } c_1=1, c_2=1, c_3=1\\
& &&\mbox{lecturer capacities: } d_1=2, d_2=1 
\end{align*}
\end{minipage}
\vspace{10pt}
\caption{A {\sc spa} instance $I$}
\label{spa_instance}
\end{figure}

A special case of {\sc spa}, where each project is offered by a unique lecturer with an infinite upper quota and zero lower quota, can be modelled as the \emph{Capacitated House Allocation problem with Ties {\sc (chat)}}. This is a variant of the well-studied \emph{House Allocation problem {\sc (ha)}} \cite{HZ79,Zho90} which involves  the allocation of a set of indivisible goods (which we call houses) to a set of applicants. In {\sc chat}, each applicant is required to rank a subset of the houses in order of preference with the houses having no preference over applicants. The applicants play the role of students and the houses play the role of projects and lecturers. As in the case of {\sc spa}, we seek to find a many-to-one matching comprising applicant-house pairs.  
Efficient algorithms for finding profile-based optimal matchings in {\sc chat} have been studied in the literature \cite{HK13,Irv06,Sng08,MM06}. The most efficient of these is the $O(R^*m\sqrt{n})$ algorithm for finding rank-maximal, greedy maximum and generous maximum matchings in {\sc chat} problems due to Huang et al.\ \cite{HK13} where $R^*$ is the maximum rank of any applicant in the matching, $m$ is the sum of all the preference list lengths and $n$ is the total number of applicants and houses. These models however fail to address the issue of load balancing among lecturers. In order to keep the assignment of students fair each lecturer will typically have a minimum (lower quota) and maximum (capacity/upper quota) number of students they are expected to supervise. These numbers may vary for different lecturers according to other administrative and academic commitments.

The {\sc chat} algorithms mentioned above are based on modelling the problem in terms of a bipartite graph with the aim of finding a matching in the graph which satisfies the stated criteria. However a more flexible approach would be to model the problem as a network with the aim of finding a flow that can be converted to a matching which satisfies the stated criteria. {\sc spa} has also been investigated in the network flow context \cite{Abr03,MU14} where a  \emph{minimum cost maximum flow algorithm} is used to find a minimum cost maximum matching and other profile-based optimal matchings. The model presented in  \cite{MU14} allows for lower quotas on lecturers and projects as well as alternative lecturers to supervise each project. By an appropriate assignment of edge weights in the network it is shown that a minimum cost maximum flow algorithm  (due to Orlin \cite{Or88a}) can find rank maximal, generous maximum and greedy maximum matchings in a {\sc spa} instance. 
This takes $O(m\log n(m + n\log n))$ time in the worst case, where $m$ and $n$ are the number of vertices and edges in the network respectively. 
In the {\sc spa} context this takes $O(m_2^2\log n_1 + m_2n_1\log^2n_1)$ time where $n_1$ is  the numbers of students and $m_2$ is the sum of all the students' preference list lengths.
However this approach involves assigning exponentially large edge weights (see, e.g., \cite[p.405]{Man13}), which may be computationally infeasible for larger problem instances due to floating point inaccuracies in dealing with such high numbers. For example given a large {\sc spa} instance involving say, $n_1=100$ students each ranking $R=10$ projects in order of preference, edge weights could potentially be of the order $n_1^R=100^{10} = 10^{20}$ (and arithmetic involving such weights could easily require more than the $15$-$17$ significant figures available in a  $64$-bit double-precision floating representation). Since the flow algorithms involve comparing these edge weights, floating point precision errors could easily cause them to fail in practice. Moreover using the standard assumption that arithmetic on numbers of magnitude $O(n_1)$ takes constant time, arithmetic on edge weights of magnitude $O(n_1^R)$ would add an additional factor of $O(R)$ onto the running time of Orlin's algorithm.

%----------------------------------------------------------------------------------------------------------------------
\subsection{Other {\sc spa} models and approaches}
\label{other-models}
The variants of \scp{spa} already discussed above have been motivated by both practical and theoretical interests.
These variants are usually distinguished by the (i) feasibility and (ii) optimality criteria specific to them.
In this section, we discuss some more \scp{spa} models found in the literature as well as other approaches that have been used to solve these problems. 
The techniques employed include \emph{Integer Programming (IP)} \cite{AB03,VSc13,Sab01, Kba13},  \cite{Sab01, Kba13}, \emph{Constraint Programming (CP)} \cite{Dye01,Tho03}, and others \cite{TH98,HSVS05,PCHJ09}. 

In \cite{Sab01}, an IP model for {\sc spa} was presented with the aim of optimising the overall satisfaction of the students and the lecturers offering the projects (i.e., minimising the overall cost on both sides).  In \cite{AB03} an IP model was presented for {\sc spa} problems involving individual and group projects. Various objective functions were also employed (often in a hierarchical manner). These include minimising the cost, balancing the work-load among lecturers,  maximising the number of students assigned and  maximising the number of first-choice assignments (w.r.t. student preferences). In \cite{VSc13} a more general IP model for {\sc spa} which allows project lower quotas was also presented. However none of these models simultaneously consider  profile-based optimality as well as upper and lower quota constraints.

%----------------------------------------------------------------------------------------------------------------------
\subsection{Our contribution}
 In Section \ref{spa_defs} we formally define the \scp{spa} model. In Section \ref{spa_greedy} we present an $O(n_1^2Rm_2)$ time algorithm for finding a greedy maximum matching given a {\sc spa} instance and prove its correctness.  The algorithm takes lecturer upper quotas into consideration.
In Section \ref{generous} we show how this algorithm can be modified in order to find a generous maximum matching. 
Section \ref{lower-q} introduces lecturer lower quotas to the \scp{spa} model and shows how our algorithm can be modified to handle this variant.
In Section \ref{spa_impl} we present results from an empirical evaluation of the algorithms described. We conclude the paper in Section \ref{ch7_conclusion} by presenting some open problems. 

%--------------------------------------------------------------------------------------------------------------------------------------------------
\section{Preliminary definitions}
\label{spa_defs}
An instance $I$ of the {\sc spa} problem consists of a set $\mathcal{S}$ of students, a set $\mathcal{P}$ of projects and a set $\mathcal{L}$ of lecturers. Each student $s_i$ ranks a set $A_i \subseteq \mathcal{P}$ of projects that she considers acceptable in order of preference. This \emph{preference list} of projects may contain ties. Each project $p_j \in \mathcal{P}$ has an upper quota $c_j$ indicating the maximum number of students that can be assigned to it. Each lecturer $l_k \in \mathcal{L}$ offers a set of projects $P_k \subseteq \mathcal{P}$ and has an upper quota $d_k^+$ indicating the maximum number of students that can be assigned to $l_k$. Unless explicitly mentioned, we assume that all lecturer lower quotas are equal to $0$. The sets $\{P_1,\dots,P_k\}$ partition $\mathcal P$. If project $p_j \in P_k$, then we denote $l_k = l(p_j)$. 

An \emph{assignment} $M$ in $I$ is a subset of $\mathcal{S} \times \mathcal{P}$ such that:
\begin{enumerate}
\item Student-project pair $(s_i, p_j) \in M$ implies $p_j \in A_i$.
\item For each student $s_i \in \mathcal{S},~|\{(s_i, p_j) \in M : p_j \in A_i\}| \leq 1$.
\end{enumerate}
If $(s_i, p_j) \in M$ we denote $M(s_i) = p_j$. For a project $p_j$, $M(p_j)$ is the set of students assigned to $p_j$ in $M$. Also if $(s_i, p_j) \in M$ and $p_j \in P_k$ we say student $s_i$ is assigned to project $p_j$ and to  lecturer $l_k$ in $M$. We denote the set of students  assigned to a lecturer $l_k$ as $M(l_k)$. A \emph{matching} in this problem is an assignment $M$ that satisfies the capacity constraints of the projects and lecturers. That is, $|M(p_j)| \leq c_j$ for all projects $p_j \in \mathcal{P}$ and $|M(l_k)| \leq d_k^+$ for all lecturers $l_k \in \mathcal{L}$.

Given a student $s_i$ and a project $p_j \in A_i$, we define  $rank(s_i, p_j)$ as $1~+ $ the number of projects that $s_i$ prefers to $p_j$. Let $R$ be the maximum rank of a project in any student's preference list.  We define the \emph{profile} $\rho(M)$ of a matching $M$ in $I$ as an $R$-tuple $(x_1, x_2, ..., x_R)$ where for each $r$ ($1\leq r\leq R$), $x_r$ is the number of students $s_i$ assigned in $M$ to a project $p_j$ such that $rank(s_i,p_j)=r$.
Let $\alpha = (x_1, x_2, ..., x_R)$ and $\sigma = (y_1, y_2, ..., y_R)$ be any two profiles. 
We define the \emph{empty profile} $O_R = (o_1, o_2, ..., o_R)$ where $o_r=0$ for all $r~(1 \leq r \leq R)$. 
We also define the \emph{negative infinity profile} $B_R^- = (b_1, b_2, ..., b_R)$ where $b_r=-\infty$ ($1\leq r\leq R$) and the \emph{positive infinity profile} $B_R^+ = (b_1, b_2, ..., b_R)$ where $b_r=\infty$ ($1\leq r\leq R$). We define the sum of two profiles $\alpha$ and $\sigma$ as $\alpha + \sigma = (x_1+y_1, x_2+y_2, ..., x_R+y_R)$. Given any $q~(1 \leq q \leq R)$, we define $\alpha + q = (x_1, ..., x_{q-1}, x_{q}+1, x_{q+1}, ..., x_R)$. We define $\alpha - q$ in a similar way.

We define the total order $\succ_L$ on profiles as follows. We say $\alpha$ \emph{left dominates} $\sigma$, denoted by $\alpha \succ_L \sigma$ if there exists some $r~(1\leq r \leq R)$ such that $x_{r'} = y_{r'}$ for $1 \leq r' < r$ and $x_r > y_r$. We define \emph{weak left domination} as follows. We say $\alpha \succeq_L \sigma$ if $\alpha=\sigma$ or $\alpha \succ_L \sigma$.
We may also define an alternative total order $\prec_R$ on profiles as follows. We say $\alpha$ \emph{right dominates} $\sigma$ ($\alpha \prec_R \sigma$) if there exists some $r~(1\leq r \leq R)$ such that $x_{r'} = y_{r'}$ for $r < r' \leq R$ and $x_r < y_r$. We also define \emph{weak right domination} as follows. We say $\alpha \preceq_R \sigma$ if $\alpha=\sigma$ or $\alpha\prec_R \sigma$.

The {\sc spa} problem can be modelled as a network flow problem. Given a {\sc spa} instance $I$, we construct a flow network $N(I) = \langle G, c \rangle$ where $G = (V, E)$ is a directed graph and $c$ is a non-negative capacity function $c : E \rightarrow \mathbb{R}^+$ defining the maximum flow allowed through each edge in $E$. The network consists of a single source vertex $v_s$ and sink vertex $v_t$ and is constructed as follows. Let $V = \{v_s, v_t\} \cup \mathcal{S} \cup \mathcal{P} \cup \mathcal{L}$ and $E = E_1 \cup E_2 \cup E_3 \cup E_4$ where $E_1=\{(v_s, s_i) : s_i \in \mathcal{S}\}$, $E_2 = \{(s_i, p_j) : s_i \in \mathcal{S}, p_j \in A_i\}$, $E_3 = \{(p_j, l_k) : p_j \in \mathcal{P}, l_k =l(p_j)\}$ and $E_4 = \{(l_k, v_t) : l_k \in \mathcal{L}\}$. We set the capacities as follows: $c(v_s, s_i) = 1$ for all $(v_s, s_i) \in E_1$, $c(s_i, p_j) = 1$ for all $(s_i, p_j) \in E_2$, $c(p_j, l_k) = c_j$ for all $(p_j, l_k) \in E_3$ and $c(l_k, v_t) = d_k^+$ for all $(l_k, v_t) \in E_4$.

We call a path $P'$ from $v_s$ to some project $p_j$ a \emph{partial augmenting path} if $P'$ can be extended  adding the edges $(p_j, l(p_j))$ and $(l(p_j), v_t)$ to form an augmenting path with respect to flow $f$. Given a partial augmenting path $P'$ from $v_s$ to $p_j$, we define the \emph{profile} of $P'$, denoted $\rho(P')$, as follows:
$$\rho(P') = O_R + \sum\{rank(s_i, p_j) : (s_i, p_j) \in P' \wedge f(s_i, p_j)=0\} ~- $$
$$\sum\{rank(s_i, p_j): (p_j, s_i) \in P' \wedge f(s_i, p_j)=1\}$$
where additions are done with respect to the $+$ and $-$ operations on profiles. Unlike the profile of a matching, the profile of an augmenting path may contain negative values. Also if $P'$ can be extended to a full augmenting path $P$ with respect to flow $f$ by adding the edges $(p_j, l(p_j))$ and $(l(p_j), v_t)$ where $v_s$ and $p_j$ are the endpoints of $P'$, then  we define the profile of $P$, denoted by $\rho(P)$, to be $\rho(P) = \rho(P')$. Multiple partial augmenting paths may exist from $v_s$ to $p_j$, thus we define the \emph{maximum profile of a partial augmenting path} from $v_s$ to $p_j$ with respect to $\succ_L$, denoted $\Phi(p_j)$, as follows:
\begin{eqnarray*}
\begin{minipage}{\linewidth}
\centering
$\Phi(p_j) = \max_{\succ_L}\{\rho(P'): P' $ is a partial augmenting path from $v_s$ to $p_j\}$. 
\end{minipage}
\end{eqnarray*}
An augmenting path $P$ is called a \emph{maximum profile augmenting path} if 
\begin{eqnarray*}
\begin{minipage}{\linewidth}
\centering
$\rho(P) = \max_{\succ_L}\{\Phi(p_j) : p_j \in \mathcal{P}\}$.
\end{minipage}
\end{eqnarray*}

Let $f$ be an integral flow in $N$. We define the matching $M(f)$ in $I$ induced by $f$ as follows: $M(f) = \{(s_i, p_j) : f(s_i, p_j)=1\}$. Clearly by construction of $N$, $M(f)$ is a matching in $I$, such that $|M(f)| = |f|$. If $f$ is a flow and $P$ is an augmenting path with respect to $f$ then $\rho(M') = \rho(M) + \rho(P)$ where $M=M(f), M'=M(f')$ and $f'$ is the flow obtained by augmenting $f$ along $P$. 
Also given a matching $M$ in $I$, we define a flow $f(M)$ in $N$ corresponding to $M$ as follows:
\begin{eqnarray*}
\centering
\begin{minipage}{400px}
$\forall~(v_s, s_i) \in E_1,$ $f(v_s, s_i) = 1$ if $s_i$ is matched in $M$ and $f(v_s, s_i) = 0$ otherwise. \\
$\forall~(s_i, p_j) \in E_2,$ $f(s_i, p_j) = 1$ if $(s_i, p_j) \in M$ and $f(s_i, p_j) = 0$ otherwise.\\
$\forall~(p_j, l_k) \in E_3,$ $f(p_j, l_k) = c'_j$ where $c'_j=|M(p_j)|$\\
$\forall~(l_k, v_t) \in E_4,$ $f(l_k, v_t) = d'_k$ where $d'_k=|M(l_k)|$
\end{minipage}
\end{eqnarray*}
We define a student $s_i$ to be \emph{exposed} if $f(v_s, s_i)=0$ meaning that there is no flow through $s_i$. Similarly we define a project $p_j$ to be \emph{exposed} if $f(p_j, l_k) < c_j$ and $f(l_k, v_t) < d_k^+$ where $l_k = l(p_j)$.

Let $M$ be a matching of size $k$ in $I$. We say that $M$ is a \emph{greedy $k$-matching} if there is no other matching $M'$ such that $|M'|=k$ and $\rho(M') \succ_L \rho(M)$. If $k$ is the size of a maximum cardinality matching in $I$, we call $M$ a \emph{greedy maximum matching} in $I$. Also we say that $M$ is a \emph{generous $k$-matching} if there is no other matching $M'$ such that $|M'| = k$ and $\rho(M') \prec_R \rho(M)$. If $k$ is the size of a maximum cardinality matching in $I$, we call $M$ a \emph{generous maximum matching} in $I$.  We also define the \emph{degree} of a matching $M$ to be the rank of one of the worst-off students matched in $M$ or $0$ if $M$ is an empty set.

%--------------------------------------------------------------------------------------------------------------------------------------------------
\section{Greedy maximum matchings in {\sc spa}}
\label{spa_greedy}
In this section we present the algorithm \sfp{Greedy-max-spa} for finding a greedy maximum matching given a {\sc spa} instance. The algorithm is based on the general Ford-Fulkerson algorithm for finding a maximum flow in a network \cite{FF62}. We obtain maximum profile augmenting paths by adopting  techniques used in the bipartite matching approach for finding a greedy maximum matching in {\sc ha} \cite{Irv06} and  {\sc chat} \cite{Sng08}. 

The \sfp{Greedy-max-spa} algorithm shown in Algorithm \ref{gen-spa} takes in a {\sc spa}  instance $I$ as input and returns a greedy maximum matching $M$ in $I$. A flow network $N(I) = \langle G, c \rangle$ is constructed as described in Section \ref{spa_defs}. Given a flow $f$ in $N(I)$ that yields a greedy $k$-matching $M(f)$ in $I$, if $k$ is not the size of a maximum flow in $N(I)$, we seek to find a maximum profile augmenting path $P$ with respect to $f$ in $N(I)$ such that the new flow $f'$ obtained by augmenting $f$ along $P$ yields a greedy $(k+1)$-matching $M(f')$ in $I$.  
Lemmas \ref{kwa1}  and \ref{kwa2} show the correctness of this approach. 
We firstly show that if $k$ is smaller than the size of a maximum flow in $N(I)$ then such a path is bound to exist. 

\begin{lemma1}
Let $I$ be an instance of {\sc spa} and let $\eta$ denote the size of a maximum matching in $I$. Let $k~(1 \leq k < \eta)$ be given and suppose that $M_k$ is a greedy $k$-matching in $I$. Let $N =N(I)$ and $f = f(M_k)$.  Then there exists an augmenting path $P$ with respect to $f$ in $N$ such that if $f'$ is the result of augmenting $f$ along $P$ then $M_{k+1} = M(f')$ is a greedy $(k+1)$-matching in $I$.
\label{kwa1}
\end{lemma1}
\begin{proof}
Let $I' = C(I)$ be a new instance of {\sc spa} obtained from $I$ as follows. Firstly we add all students in $I$ to $I'$.  Next, for every project $p_j \in \mathcal{P}$, we add $c_j$ clones $p_j^1, p_j^2, ..., p_j^{c_j}$ to $I'$ each of capacity $1$. We then add all lecturers in $I$ to $I'$. If $p_j \in A_i$ in $I$, we add $(s_i, p_j^r)$ to $I'$ for all $r~(1 \leq r \leq c_j)$. If $p_j \in P_k$ is in $I$, we add $(p_j^r, l_k)$ to $I'$  for all $r~(1 \leq r \leq c_j)$. Also if $rank(s_i, p_j) = t$, we set $rank(s_i, p_j^r)=t$ for all $r~(1 \leq r \leq c_j)$. Let $G'$ be the underlying graph in $I'$ involving only the student and project clones. With respect to the matching $M_k = M(f)$, we construct a cloned matching $C(M_k)$ in $I'$ as follows. If project $p_j$ is assigned $x_j$ students $s_{q,1}, s_{q,2}, ..., s_{q,x_j}$ in $M_k$ we add $(s_{q,r}, p_j^r)$ to $C(M_k)$ for all $1 \leq r \leq x_j$. Hence $C(M_k)$ is a greedy $k$-matching in $I'$. 

Let $M'_{k+1}$ be a greedy $(k+1)$-matching in $I$ (this exists because $k < \eta$). Then $C(M'_{k+1})$ is a greedy $(k+1)$-matching in $I'$.  Let $X = C(M_k) \oplus C(M'_{k+1})$. Then each connected component of $X$ is either (i) an alternating cycle, (ii) an even-length alternating path or (iii) an odd-length alternating path in $G'$ (with no restrictions on which matching the end edges belong to). The aim is to show that, by eliminating a subset of $X$, we are left with a set of connected components which can be transformed into a single augmenting path with respect to $f(C(M_k))$ in $N(I')$ and subsequently a single augmenting path with respect to $f(M_k)$ in $N(I)$.\\

\noindent {\bf Eliminating connected components of $X$:} Suppose $D \subseteq X$ is a type (i) connected component of  $X$ or a type (ii) connected component of $X$ whose end vertices are students (we may call this a type (ii)(a) component). Suppose also that $\rho(D\cap C(M'_{k+1})) \succ_L \rho(D \cap C(M_k))$. A new matching $C(M'_k)$ in $G'$ of cardinality $k$ can be created from $C(M_k)$ by replacing all the $C(M_k)$-edges in $D$ with the $C(M'_{k+1})$-edges in $D$ (i.e. by augmenting $C(M_k)$ along $D$). Since the upper quota constraints of the lecturers involved are not violated after creating $C(M'_k)$ from $C(M_k)$, it follows that $C(M'_k)$ is also a valid {\sc spa} matching in $I'$. Moreover $\rho(C(M_k')) \succ_L \rho(C(M_k))$ which is a contradiction to the fact that $C(M_k)$ is a greedy $k$-matching in $I'$. A similar contradiction (to the fact that $C(M'_{k+1})$ is a greedy $(k+1)$- matching in $I'$) exists if we assume $\rho(D \cap C(M_k)) \succ_L \rho(D\cap C(M'_{k+1}))$. Thus $\rho(D\cap C(M'_{k+1})) = \rho(D \cap C(M_k))$. 
 
Form the argument above, no type (i) or type (ii)(a) connected component of $X$ contributes to a change in the size or profile as we augment from $C(M_k)$ to $C(M'_{k+1})$ or vice versa. In fact, this is true for any even-length connected component of $X$ which does not cause lecturer upper quota constraints to be violated as we augment from $C(M_k)$ to $C(M'_{k+1})$ or vice versa. The claim can further be extended to certain groups of connected components which, when considered together, (i) have equal numbers of $C(M_k)$ and  $C(M'_{k+1})$ edges and (ii) do not cause lecturer upper quota constraints to be violated as we augment from $C(M_k)$ to $C(M'_{k+1})$ or vice versa. In all these cases, it is possible to \emph{eliminate} such components (or groups of components) from consideration. Using the above reasoning, we begin by eliminating all type (i) and type (ii)(a) connected components of $X$.

Let $\mathcal{D}$ be the union of all the edges in type (i) and type (ii)(a) connected components of $X$. Let $X' = X \backslash \mathcal{D}$. Then it follows that $X' = C(M_k) \oplus C(M''_{k+1})$ for some greedy $(k+1)$-matching $C(M''_{k+1})$ in $I'$ which can be constructed by augmenting $C(M'_{k+1})$ along all type (i) and type (ii)(a) components of $X$. Thus $X'$ contains
(1) even-length alternating paths whose end vertices are project clones (we call these type (ii)(b) paths), 
(2) odd-length alternating paths whose end edges are in $C(M_k)$ (we call these type (iii)(a)  paths) and 
(3) odd-length  alternating paths whose end edges are in $C(M''_{k+1})$ (we call these type (iii)(b) paths). 
Although these alternating paths are vertex disjoint, there are special cases where two alternating paths in $X'$ may be \emph{joined together} by pairing their end project clone vertices.\\

\noindent {\bf Joining alternating paths:}
Consider some lecturer $l_q$ and project $p_j \in P_q$. We extend the notation $l(p_j)$ to include all clones of $p_j$ (i.e. $l(p_j^r)=l_q$ for all $r~(1 \leq r \leq c_j)$).  Let
$$X_q = \{(s_i, p_j^r) \in C(M_k): l_q = l(p_j^r) \wedge p_j^r \mbox{ is unmatched in } C(M''_{k+1})\} \mbox{  and } x_q = |X_q|$$
Thus $X_q$ is the set of end edges incident to project clones belonging to a subset of the type (ii)(b) and type (iii)(a) paths in $X'$. Let
$$Y_q = \{(s_i, p_j^r) \in C(M''_{k+1}): l_q = l(p_j^r) \wedge p_j^r \mbox{  is unmatched in } C(M_k)\} \mbox{  and  } y_q = |Y_q|$$
Thus $Y_q$ is the set of end edges incident to project clones belonging to a subset of the type (ii)(b) and type (iii)(b) paths in $X'$.  Also let
$$Z_q = \{p_j^r: l_q = l(p_j^r) \wedge p_j^r \mbox{  is matched in } C(M''_{k+1}) \wedge p_j^r \mbox{ is matched in } C(M_k)\} \mbox{   and  } z_q = |Z_q|$$
Thus $d_q = v_q+x_q+z_q$ and $d_q = v_q'+y_q+z_q$ where $v_q$ and $v_q'$ are the number of unassigned positions that $l_q$ has in $C(M_k)$ and $C(M_{k+1}'')$ respectively. 

Note that $v_q \geq y_q$ if and only if $v_q' \geq x_q$.  If $v_q \geq y_q$ then all the paths with end edges in $Y_q$ can be considered as valid alternating paths in $C(M_k)$ (i.e. if they are used to augment $C(M_k)$, $l_q$'s upper quota will not be violated in the resulting matching). Since $v_q' \geq x_q$ then all the paths with end edges in $X_q$ can be considered as valid alternating paths in $C(M_{k+1}'')$ (i.e. if they are used to augment $C(M_{k+1}'')$,  $l_q$'s upper quota will not be violated in the resulting matching). 

On the other hand, assume $y_q > v_q$. Then $x_q > v_q'$. Let $Y_q' \subseteq Y_q$ be an arbitrary subset of $Y_q$ of size $v_q$ and let $X_q' \subseteq X_q$ be an arbitrary subset of $X_q$ of size $v_q'$. Thus all paths with end edges in $X_q'$ and $Y_q'$ can be considered as valid alternating paths in  $C(M_{k+1}'')$ and  $C(M_{k})$ respectively.
Also $|Y_q \backslash Y_q'| ~=~ y_q -v_q ~=~ |X_q \backslash X_q'| ~=~ x_q -v_q'$. We can thus form a $1-1$ correspondence between the edges in $|Y_q \backslash Y_q'|$ and those in $|X_q \backslash X_q'|$. Let $(s_i, p_j^r) \in Y_q \backslash Y_q'$ and  $(s_{i'}, p_{j'}^{r'}) \in X_q \backslash X_q'$ be the end edges of two alternating paths in $X'$. The paths can be joined together by \emph{pairing} the clones of both end projects  thus forming a \emph{project pair}  $(p_j^r, p_{j'}^{r'})$ at $l_q$. These project pairs can be formed from all edges in $Y_q \backslash Y_q'$ and $X_q \backslash X_q'$.

In the cases where project pairs are formed, the resulting path (which we call a \emph{compound path}) may be regarded as a single path along which $C(M_k)$ or $C(M''_{k+1})$ may be augmented. 
In some cases, the two projects being paired may be end vertices of a single (or compound) alternating path. Thus pairing them together will form a cycle. Since the cycle is of even length and the lecturer's upper quota will not be violated if it is used to augment  $C(M_k)$ or $C(M_{k+1}'')$ it can be eliminated right away. 
For each lecturer $l_q \in \mathcal{L}$, once the pairings between alternating paths in $Y_q \backslash Y_q'$ and $X_q \backslash X_q'$ have been carried out (where applicable) and any formed cycles have been eliminated, we are left with a set of single or compound alternating paths of the following types (for simplicity we call all remaining alternating paths \emph{compound paths} even though they may consist of only one path). 

\begin{figure}[t]
\centering
\small
\begin{minipage}[b]{0.3\linewidth}
\begin{tikzpicture}
\tikzstyle{every_node}=[draw,circle,fill=gray,minimum size=10pt, inner sep=0.3pt]
\tikzstyle{text} = [draw,circle,fill=gray,minimum size=7pt]

\node[draw] at (1,7) {compound type (ii)(a) path};

\draw (0,6) node[every_node] (s1) [label=left:$s_1$]{};
\draw (2,6) node[every_node] (p1) [label=right:$p_1$]{};
\path [->, decoration={zigzag,segment length=4,amplitude=1.9},font=\scriptsize, line join=round]  (s1) edge[decorate] (p1);                                      
\draw (0,5) node[every_node] (s2) [label=left:$s_2$]{};
\draw (2,5) node[every_node] (p2) [label=right:$p_2$]{};
\path [->, decoration={zigzag,segment length=4,amplitude=1.9},font=\scriptsize, line join=round] (s2) edge[decorate] (p2);
\path (s2) edge (p1);  

\draw (1,4) -- (1,4.5)[dashed];
 
\draw (2,3.5) node[every_node] (p3) [label=right:$p_3$]{};
\draw (0,3) node[every_node] (s3) [label=left:$s_3$]{};
\draw (2,2.5) node[every_node] (p4) [label=right:$p_4$]{};
\path [->, decoration={zigzag,segment length=4,amplitude=1.9},font=\scriptsize, line join=round]  (s3) edge[decorate] (p4); 
\path (s3) edge (p3);  

\draw (1,1.5) -- (1,2)[dashed];
 
\draw (0,1) node[every_node] (s5) [label=left:$s_5$]{};
\draw (2,1) node[every_node] (p5) [label=right:$p_5$]{};
\path   (s5) edge (p5);                                      
\draw (0,0) node[every_node] (s6) [label=left:$s_6$]{};
\draw (2,0) node[every_node] (p6) [label=right:$p_6$]{};
\path  (s6) edge (p6);
\path [->, decoration={zigzag,segment length=4,amplitude=1.9},font=\scriptsize, line join=round] (s5) edge [decorate] (p6);    
\node[draw=none] at (1,-1) {(a)};
\end{tikzpicture}
%\caption{Compound type (ii)(a) path}
\end{minipage}
\hspace{0.5cm}
\begin{minipage}[b]{0.3\linewidth}
\begin{tikzpicture}
\tikzstyle{every_node}=[draw,circle,fill=gray,minimum size=10pt, inner sep=0.3pt]
\tikzstyle{text} = [draw,circle,fill=gray,minimum size=7pt]

\node[draw] at (1,7) {compound type (iii)(a) path};

\draw (0,6) node[every_node] (s1) [label=left:$s_1$]{};
\draw (2,6) node[every_node] (p1) [label=right:$p_1$]{};
\path [->, decoration={zigzag,segment length=4,amplitude=1.9},font=\scriptsize, line join=round]  (s1) edge[decorate] (p1);                                      
\draw (0,5) node[every_node] (s2) [label=left:$s_2$]{};
\draw (2,5) node[every_node] (p2) [label=right:$p_2$]{};
\path [->, decoration={zigzag,segment length=4,amplitude=1.9},font=\scriptsize, line join=round] (s2) edge[decorate] (p2);
\path (s2) edge (p1);  

\draw (1,4) -- (1,4.5)[dashed];
 
\draw (2,3.5) node[every_node] (p3) [label=right:$p_3$]{};
\draw (0,3) node[every_node] (s3) [label=left:$s_3$]{};
\draw (2,2.5) node[every_node] (p4) [label=right:$p_4$]{};
\path [->, decoration={zigzag,segment length=4,amplitude=1.9},font=\scriptsize, line join=round]  (s3) edge[decorate] (p4); 
\path (s3) edge (p3);  

\draw (1,1.5) -- (1,2)[dashed];
\node[draw=none] at (1,-1) {(b)};
\end{tikzpicture}

\end{minipage}
\hspace{0.5cm}
\begin{minipage}[b]{0.3\linewidth}
\begin{tikzpicture}
\tikzstyle{every_node}=[draw,circle,fill=gray,minimum size=10pt, inner sep=0.3pt]
\tikzstyle{text} = [draw,circle,fill=gray,minimum size=7pt]

\node[draw] at (1,7) {compound type (iii)(b) path};

\draw (0,6) node[every_node] (s1) [label=left:$s_1$]{};
\draw (2,6) node[every_node] (p1) [label=right:$p_1$]{};
\path (s1) edge (p1);                                      
\draw (0,5) node[every_node] (s2) [label=left:$s_2$]{};
\draw (2,5) node[every_node] (p2) [label=right:$p_2$]{};
\path  (s2) edge (p2);
\path [->, decoration={zigzag,segment length=4,amplitude=1.9},font=\scriptsize, line join=round] (s2) edge[decorate] (p1);  

\draw (1,4) -- (1,4.5)[dashed];
 
\draw (2,3.5) node[every_node] (p3) [label=right:$p_3$]{};
\draw (0,3) node[every_node] (s3) [label=left:$s_3$]{};
\draw (2,2.5) node[every_node] (p4) [label=right:$p_4$]{};
\path [->, decoration={zigzag,segment length=4,amplitude=1.9},font=\scriptsize, line join=round]  (s3) edge[decorate] (p3); 
\path (s3) edge (p4);  

\draw (1,1.5) -- (1,2)[dashed];
\node[draw=none] at (1,-1) {(c)};
\end{tikzpicture}

\end{minipage}
\begin{tikzpicture}
\draw (4,0) node (x0){};
\draw (5,0) node (x) [label=right:$ \in C(M_{k+1}'')$]{};
\path  (x0) edge (x); 
\draw (1,0) node (y) [label=right:$ \in C(M_k)$]{};
\draw (0,0) node (y0){};
\path [->, decoration={zigzag,segment length=4,amplitude=1.9},font=\scriptsize, line join=round] (y0) edge[decorate] (y); 
\end{tikzpicture}
\caption{Some types of compound path in $X'$}
\label{cp_paths}
\end{figure}
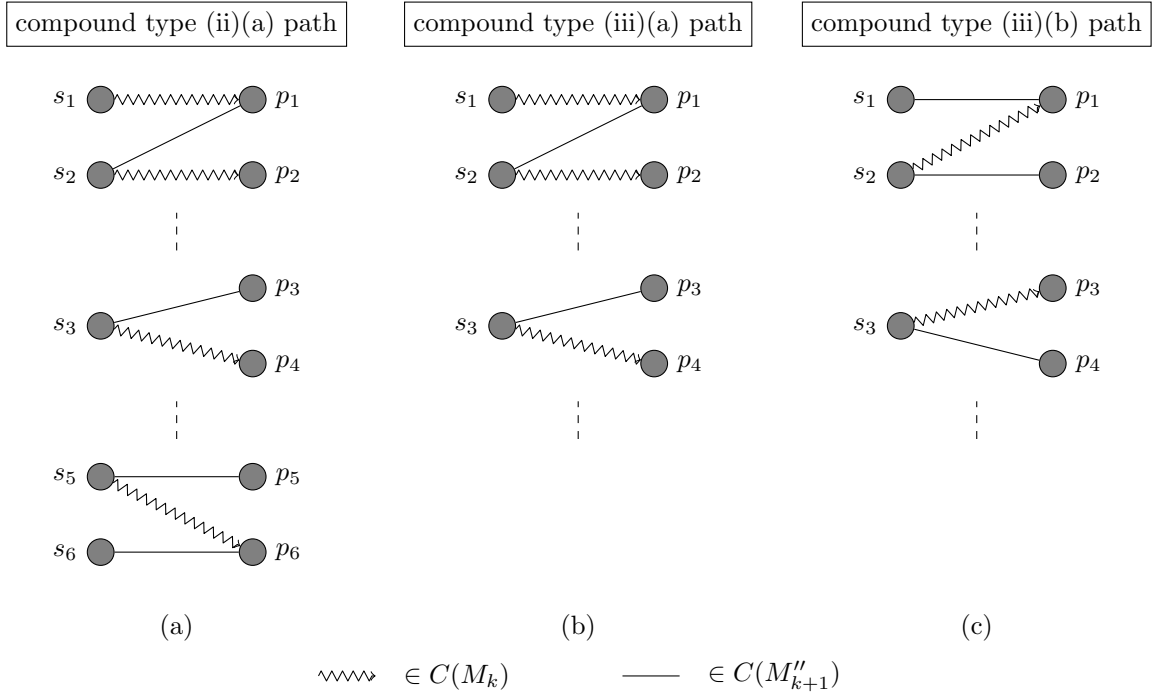

\begin{enumerate}
\item A \emph{compound type (ii)(a) path} - a compound path with an even number of edges with both end vertices being students. This path will contain a type (iii)(a) path at one end, and a type (iii)(b) path at the other end with zero or more type (ii)(b) paths in between (See Figure \ref{cp_paths}(a)). Such a path can be eliminated from consideration. 
\item A \emph{compound type (ii)(b) path} - a compound path with an even number of edges with both end vertices being project clones. This path will contain one or more type (ii)(b) paths joined together. Such a path can also be eliminated from consideration as its end edges are incident to exposed project clones. 
\item A \emph{compound type (iii)(a) path} - a compound path with an odd number of edges with both end edges being matched in $C(M_k)$. This path will contain a type (iii)(a) path at one end with zero or more type (ii)(b) paths joined to it (See Figure \ref{cp_paths}(b)). We will consider these paths for elimination later in this proof.
\item A \emph{compound type (iii)(b) path} - a compound path with an odd number of edges with both end edges being matched in $C(M_{k+1}'')$. This path will contain a type (iii)(b) path at one end with zero or more type (ii)(b) paths joined to it (See Figure \ref{cp_paths}(c)). We will consider these paths for elimination later in this proof.
\end{enumerate}

\noindent {\bf Eliminating compound paths:} At this stage we are left with only compound type (iii)(a) and compound  type (iii)(b) paths in $X'$. These paths, if considered independently decrease and increase the size of $C(M_k)$ by $1$ respectively. Since $|C(M''_{k+1})| = |C(M_k)|+1$ then there are $q$ type (iii)(a) paths and $(q+1)$ type (iii)(b) paths. Consider some compound type (iii)(b) path $D'$ and some compound type (iii)(a) path $D''$. Then we can consider the combined effect of augmenting $C(M_k)$ or $C(M''_{k+1})$ along $D' \cup D''$.  Suppose that $\rho((D' \cup D'') \cap C(M''_{k+1}))  \succ_L  \rho((D' \cup D'')\cap C(M_k))$. A new matching $C(M''_k)$ in $G'$ of cardinality $k$ can be created by augmenting $C(M_k)$ along $D' \cup D''$. Since the upper quota constraints on the lecturers involved are not violated after creating $C(M''_k)$ from $C(M_k)$, then $C(M''_k)$ is also a valid {\sc spa} matching in $I'$. Thus $\rho(C(M''_k)) \succ_L \rho(C(M_k))$ which is a contradiction to the fact that $C(M_k)$ is a greedy $k$-matching in $I'$. A similar contradiction (to the fact that $C(M''_{k+1})$ is a greedy $(k+1)$-matching in $I'$) exists if we assume $\rho((D' \cup D'') \cap C(M_k)) \succ_L \rho((D' \cup D'') \cap C(M''_{k+1}))$. Thus $\rho((D' \cup D'') \cap C(M''_{k+1})) = \rho((D' \cup D'') \cap C(M_k))$. It follows that, considering $D'$ and $D''$ together, the size and profile of the matching is unaffected as augment from $C(M_k)$ to $C(M''_{k+1})$ or vice versa and so both $D'$ and $D''$ can be eliminated from consideration.\\

\noindent {\bf Generating an augmenting path in $N(I)$:}
Once all these eliminations have been done, since $|C(M''_{k+1})| = |C(M_{k})| + 1$ it is easy to see that there remains only one path $P'$  left in $X'$ which is a compound type (iii)(b) path.
The path $P'$ can then be transformed to a component $D$ in $G(I)$ (where $G(I)$ is basically the undirected counterpart of $N(I)$ without capacities) by replacing all the project clones $p_j^r~(1 \leq r \leq c_j)$ in $P'$ with the original project $p_j$ and, for every joined pair of project clones ($p_j^r, p_{j'}^{r'}$), adding the lecturer $l(p_j^r) = l(p_{j'}^{r'})$ in between them. Thus a project may now appear more than once in $D$. A lecturer may also appear more than once in $D$. 

Consider some project $p_j \in D$ that appears more than once. Then let $P'' \subset P'$ be the path consisting of edges between the first and last occurrence of the $p_j$ clones in $P'$ ($P''$ corresponds to a collection of cycles belonging to $D$ in $G(I)$ involving $p_j$). Thus $P''$ is of even length and both end projects of $P''$ are clones of the same project. Augmenting $C(M_k)$ or $C(M''_{k+1})$ along $P''$ will not violate the lecturer upper quota constraints or affect the size or profile of the matching obtained (again using the same arguments presented above). Thus $P''$ can be eliminated from consideration. Although this potentially breaks $P'$ into two separate paths in $G(I')$ it still remains connected in $G(I)$. 
Similarly consider some lecturer $l_k \in D$ that appears more than once. Then let $P''' \subset P'$ be the path consisting of edges between the first and last occurrence of the $l_k$ clones in $P'$ ($P'''$ corresponds to a collection of type (ii)(b) paths with project clones offered by $l_k$). 
Thus augmenting $C(M_k)$ or $C(M''_{k+1})$ along $P'''$ will not violate the lecturer upper quota constraints or affect the size or profile of the matching obtained (again using the same arguments presented above). Thus $P'''$ can be eliminated from consideration. Doing the above steps continually for all projects and lecturers that occur more than once in $D$ eventually yields a valid path in $G(I)$ in which all nodes are visited only once.

Finally we describe how the path $D$ in $G(I)$, obtained after removing duplicate projects and lecturers, can be transformed to an augmenting path $P$ in $N(I)$ (i.e. we establish the direction of flow from $v_s$ to $v_t$ through $P$ in $N(I)$). Firstly we add the edge $(v_s, s_i)$ to $P$ where $s_i$ is the exposed student in $D$. Next for every edge $(s_{i'}, p_{j'}) \in M''_{k+1} \cap D$ we add a forward edge $(s_{i'}, p_{j'})$ to $P$. Also for every edge $(s_{i''}, p_{j''}) \in M_{k} \cap D$ we add a backward edge $(p_{j''},s_{i''})$ to $P$. Finally we add the edges $(p_j, l(p_j))$ and $(l(p_j), v_t)$ to $P$ where $p_j^r$ is the end project vertex in $D$. Thus $P$ is an augmenting path with respect to $f = f(M_k)$ in $N(I)$ such that if $f'$ is the flow obtained when $f$ is augmented along $P$ then $M(f')$ is a greedy $(k+1)$-matching in $N(I)$.
\end{proof}

\begin{lemma1}
Let $f$ be a flow in $N$ and let $M_k = M(f)$. Suppose that $M_k$ is a greedy $k$-matching. Let $P$ be a maximum profile augmenting path with respect to $f$. Let $f'$ be the flow obtained by augmenting $f$ along $P$. Now let $M_{k+1} = M(f')$. Then $M_{k+1}$ is a greedy $(k+1)$-matching. 
\label{kwa2}
\end{lemma1}
\begin{proof}
Suppose for a contradiction that $M_{k+1}$ is not a greedy $(k+1)$-matching. By Lemma \ref{kwa1}, there exists an augmenting path $P'$ with respect to $f$ such that if $f'$ is the result of augmenting $f$ along $P'$ then $M_{k+1}' = M(f')$ is a greedy $(k+1)$-matching. 
Hence $\rho(M_{k+1}') \succ_L \rho(M_{k+1})$. Since $\rho(M_{k+1}') = \rho(M) + \rho(P')$ and $\rho(M_{k+1}) = \rho(M) + \rho(P)$, it follows that $\rho(P') \succ_L \rho(P)$, a contradiction to the assumption that $P$ is a maximum profile augmenting path.
\end{proof}

\begin{algorithm}[t]
\small
\caption{\sfp{Greedy-max-spa}}
\label{gen-spa}
\begin{algorithmic}[1]
	\REQUIRE {\sc spa} instance $I$;
	\ENSURE return matching $M$;
	\STATE define flow network $N(I) = \langle G, c \rangle$;
	\STATE define empty flow $f$; \label{empty}
	\LOOP
	\STATE $P = $ \sfp{Get-max-aug}$(N(I), f)$; \label{restart} 
	\IF {$P \neq null$}
		\STATE augment $f$ along $P$;
	\ELSE
		\STATE return $M(f)$;
	\ENDIF
	\ENDLOOP
\end{algorithmic}
\normalsize
\end{algorithm}

The \sfp{Get-max-aug} algorithm shown in Algorithm \ref{get_auth} accepts a flow network $N(I)$ and flow $f$ as input and finds an augmenting path of maximum profile relative to $f$ or reports that none exists. The latter case implies that $M(f)$ is already a greedy maximum matching. The method consists of three phases: an initialisation phase (lines \ref{alg_init} -\ref{alg_init_end}), the main phase which is a loop containing two other loops (lines \ref{alg_main} - \ref{alg_main_end}) and a final phase (lines \ref{alg_final} - \ref{alg_final_end}) where the augmenting path is generated and returned.  

For each project $p_j$ the \sfp{Get-max-aug} method maintains a variable $\rho(p_j)$ describing the profile of a partial augmenting path $P'$ from some exposed student to $p_j$. It also maintains, for every project $p_j \in \mathcal{P}$, a pointer $pred(p_j)$ to the student or lecturer preceding $p_j$ in $P'$. For every lecturer $l_k \in \mathcal{L}$ a pointer $pred(l_k)$ is also used to refer to any project preceding $l_k$ in $P'$.  Thus the final augmenting path produced will pass through each lecturer or project at most once. The initialisation phase of the method involves setting all $pred$ pointers to {\bf null} and $\rho$ profiles to $B_R^-$. Next, the method seeks to find, for each project $p_j$, a partial augmenting path $((v_s, s_i), (s_i, p_j))$ from the source, through an exposed student $s_i$ to $p_j$ should one exist. In the presence of multiple paths satisfying this criterion, the path with the best profile (w.r.t. $\succ_L$) is selected. The variables $pred(p_j)$ and $\rho(p_j)$ are updated accordingly. Thus at the end of this phase $\rho(p_j)$ indicates the maximum profile of an augmenting path of length $2$ via some exposed student to $p_j$ should one exist. If such a path does not exist then $\rho(p_j)$ and $pred(p_j)$ remain $B_R^-$ and {\bf null} respectively.

In the main phase, the algorithm then runs $|f|$ iterations, at each stage attempting to increase the quality (w.r.t. $\succ_L$)  of the augmenting paths described by the $\rho$ profiles. Each iteration runs two loops. Each loop identifies cases where the flow through one edge in the network can be reduced in order to allow the flow through another to be increased while improving the profile of the projects involved. In both loops, the decision on whether to switch the flow between candidate edges is made based on an edge relaxation operation similar to that used in the Bellman-Ford algorithm for solving the single source shortest path problem in which edge weights may be negative. In the first loop, we seek to evaluate the gain that may be derived from switching the flow through a student from one project to another. Given an edge $(s_i, p_k)$ with a flow of $1$ in $f$ and edge $(s_i, p_j)$ with no flow in $f$, we define $\sigma$ to be the resulting profile of $p_j$ if the partial augmenting path ending at $p_k$ is to be extended (via $s_i$) to $p_j$. Thus $\sigma$ will become the new value of $\rho(p_j)$ should this extension take place. If $\sigma \succ_L \rho(p_j)$ (i.e. if the proposed profile is better than the current one), we extend the augmenting path to $p_j$ and update $\rho(p_j) = \sigma$ and $pred(p_j) = s_i$. 

In the second loop, we seek to evaluate the gain that may be derived from switching flow to some lecturer from one project to another. Given a lecturer $l_k$, let $P'_k \subseteq P_k$ be the set of projects offered by $l_k$ with positive outgoing flow and $P''_k \subseteq P_k$ be the set of projects offered by $l_k$ that are undersubscribed in $M(f)$. Then we seek to determine if an improvement can be obtained by switching a unit of flow from some project $p_j \in P'_k$ to some other project $p_m \in P''_k$. This is achieved by comparing the $\rho(p_j)$ and $\rho(p_m)$ profiles and updating $\rho(p_j) = \rho(p_m)$, $pred(p_j) = l_k$ and $pred(l_k) = p_m$ if $\rho(p_m) \succ_L\rho(p_j)$ where $\rho(p_m)$ represents the profile of a partial augmenting path that does not already pass through $l_k$ (i.e., $pred(p_m) \neq l_k$). This means that the partial augmenting path ending at $p_m$ can be extended further (via $l_k$) to $p_j$ while improving its profile. The intuition is that, after augmenting along such a path, $p_m$ gains an extra student while $p_j$ loses one.

During the final phase, we iterate through all exposed projects and find the one with the largest profile with respect to $\succ_L$ (say $p_q$). An augmenting path is then constructed through the network using the $pred$ values of the projects and lecturers and the matched edges in $M(f)$ starting from $p_q$. The generated path is returned to the calling algorithm. If no exposed project exists, the method returns {\bf null}. We next show that \sfp{Get-max-aug} method produces such a maximum profile augmenting path in $N$ with respect to $f$ should one exist.

\begin{algorithm}[]
\small
\caption{\sfp{Get-max-aug} (method for \sfp{Greedy-max-spa})}
\label{get_auth}
\begin{algorithmic}[1]
\REQUIRE flow network $N(I) = \langle G, c \rangle$ where $G=(V,E)$, flow $f$ where $M(f)$ is a greedy $|f|$-matching;
\STATE /*  initialisation */ \label{alg_init}
\FOR {project $p_j \in \mathcal{P}$}
	\STATE $\rho(p_j) = B_R^-$; \label{set_neg_profile}
	\STATE $pred(p_j) = {\bf null}$;
	\FORALL{ exposed student $s_i \in S$ such that $p_j \in A_i$}
		\STATE $\sigma = O_R + rank(s_i, p_j)$;
		\IF {$\sigma \succ_L \rho(p_j)$}
			\STATE $\rho(p_j) = \sigma$; 
			\STATE $pred(p_j) = s_i$;
		\ENDIF
	\ENDFOR
\ENDFOR
\FOR {lecturer $l_k \in \mathcal{L}$}
	\STATE $pred(l_k) = {\bf null}$;
\ENDFOR \label{alg_init_end}
\STATE /*  main phase */ \label{alg_main}
\FOR {$1...|f|$}
\STATE /*  first loop */
\FORALL {$(s_i, p_j) \in E$ where $f(s_i, p_j) = 0$ and $f(s_i, p_k)=1$ for some $p_k \in A_i$}
	\STATE $\sigma = \rho(p_k) - rank(s_i, p_k) + rank(s_i, p_j)$;
	\IF {$\sigma \succ_L \rho(p_j)$}
		\STATE $\rho(p_j) = \sigma$;  $pred(p_j) = s_i$;
	\ENDIF
\ENDFOR
\STATE /*  second loop */ \label{sec_loop}
	\FORALL {lecturer $l_k \in \mathcal{L}$}
		\STATE $\sigma = B_R^-$;
		\STATE $p_z = {\bf null}$;
		\FORALL {project $p_m \in P_k$ such that $l(p_m) = l_k \wedge f(p_m, l_k) < c_m$} \label{extra_condition}
			\IF {$\rho(p_m) \succ_L \sigma$}
					\STATE $\sigma = \rho(p_m)$;
					\STATE $p_z = p_m$;
			\ENDIF
		\ENDFOR
		\IF {$p_z \neq {\bf null}$}
		\FORALL {project $p_j \in P_k$ such that $l(p_j) = l_k \wedge f(p_j, l_k) > 0 \wedge p_j \neq p_z$}
				\STATE $\rho(p_j) = \sigma$;
				\STATE $pred(p_j) = l_k$;
				\STATE $pred(l_k) = p_z$;
		\ENDFOR
		\ENDIF
	\ENDFOR
\ENDFOR \label{alg_main_end}
\algstore{myalg}
\end{algorithmic}
\normalsize
\end{algorithm}

\begin{algorithm}[]
\small
\begin{algorithmic}[1]
\algrestore{myalg}
\STATE /*  final phase */ \label{alg_final}
\STATE $\rho = \max_{\succ_L}(\{B_R^-\} \cup \{\rho(p_j) : p_j \in P$ is exposed$\})$; \label{first_max}
\IF{$\rho \succ_L B_R^-$}
	\STATE $p_q = $ arg $ \max_{\succ_L}(\{B_R^-\} \cup \{\rho(p_j) : p_j \in P$ is exposed$\})$; \label{second_max}
	\STATE $Q =$ path obtained by following $pred$ values  and matched edges in $M(f)$ from $p_q$ to an exposed student;
	\STATE return $\langle v_s \rangle$ {\tt++} $reverse(Q)$ {\tt++} $\langle l(p_q), v_t \rangle$;  /*{\tt++} denotes concatenation*/
\ELSE
	\STATE return {\bf null};
\ENDIF \label{alg_final_end}
\end{algorithmic}
\normalsize
\end{algorithm}

\begin{lemma1}
Given a \scp{spa} instance $I$, let $f$ be a flow in $N= N(I)$ where $k = |f|$ is not the size of a maximum matching in $I$ and $M(f)$ is a greedy $k$-matching in $I$. Algorithm \sfp{Get-max-aug} finds a maximum profile augmenting path in $N$ with respect to $f$.
\label{kwa3}
\end{lemma1}
\begin{proof}
Consider some project $p_j$ in $\mathcal{P}$. 
For any $q~(0 \leq q \leq k)$ and for any $r~(0 \leq r \leq k)$, we define $\Phi_{2q+1, 2r}(p_j)$ to be the maximum profile of any partial augmenting path with respect to $f$ in $N$ that starts at an exposed student, ends at $p_j$, and involves at most $2q+1$ student-project edges and at most $2r$ project-lecturer edges. We represent the length of such a path using the pair $(2q+1, 2r)$.
Thus $\Phi_{2k+1, 2k}(p_j)$ gives the maximum profile of any partial augmenting path starting at an exposed student and ending at $p_j$.
If such a path does not exist then $\Phi_{2k+1, 2k}(p_j) = B_R^-$. 
Firstly we seek to show that after $q$ iterations of the main loop of \sfp{Get-max-aug} where $0 \leq q \leq k$, $\rho_{q}(p_j) \succeq_L \Phi_{2q+1,2q}(p_j)$ for every project $p_j \in \mathcal{P}$ where $\rho_{q}(p_j)$ is the profile computed at $p_j$ after $q$ iterations of the main loop.

We prove this inductively. For the base case, let $q=0$. Then  $\Phi_{1,0}(p_m)$ is the maximum profile of any partial augmenting path of length $(1,0)$ from an exposed student to project $p_m$. Hence, from the initialisation phase of \sfp{Get-max-aug}, $\rho_0(p_m) = \Phi_{1,0}(p_m)$ and thus $\rho_0(p_m) \succeq_L \Phi_{1,0}(p_m)$. 
For the inductive step, assume $1 \leq q \leq k$ and that the claim is true after the $(q-1)^{th}$ iteration (i.e. $\rho_{q-1}(p_m) \succeq_L \Phi_{2q-1, 2q-2}(p_m)$ for any $p_m \in \mathcal{P}$). We will show that the claim is true for the $q^{th}$ iteration (i.e. $\rho_{q}(p_m) \succeq_L \Phi_{2q+1,2q}(p_m)$). 

For each project $p_m \in \mathcal{P}$ let $S'_m = \{s_i \in \mathcal{S} : (s_i, p_m) \in E \wedge f(s_i, p_m) = 0 \}$ and for each lecturer $l_k \in \mathcal{L}$ let $P'_k = \{p_m \in \mathcal{P}: l_k = l(p_m) \wedge f(p_m, l_k)<c_m \}$. 
For each iteration of the main loop, we perform a relaxation step involving some student-project pair $(s_i, p_m)$ where $s_i \in S'_m$ and/or a relaxation step involving some project-lecturer pair $(p_m, l_k)$ where $p_m \in P'_k$.  
Consider some project $p_m$. 
If there does not exist a partial augmenting path from an exposed student to $p_m$, of length $\leq (2q+1, 2q-2)$ and with a better profile than $\Phi_{2q-1, 2q-2}(p_m)$, then $\Phi_{2q+1,2q-2}(p_m) = \Phi_{2q-1,2q-2}(p_m)$. 
Otherwise there exists a partial augmenting path from an exposed student to $p_m$ of length at least $(2q+1, 2q-2)$ with a better profile than $\Phi_{2q-1,2q-2}(p_m)$. 
Such a path must contain a partial augmenting path from an exposed student to some project $p_{m'}$ such that:
\begin{eqnarray*}
\centering
\begin{minipage}{0.9\linewidth}
$\Phi_{2q+1, 2q-2}(p_m) = \Phi_{2q-1,2q-2}(p_{m'}) + rank(s_i, p_m) - rank(s_i, p_{m'})$.
\end{minipage}
\end{eqnarray*}
where $ s_i \in S_m'$ and $f(s_i, p_{m'})=1$. Thus we  note the following identity involving  $\Phi_{2q+1,2q-2}(p_m)$:
\begin{eqnarray}
\label{lbl_a}
\centering
\begin{minipage}{0.9\linewidth}
$\Phi_{2q+1,2q-2}(p_m) = \max_{\succ_L}\{\Phi_{2q-1,2q-2}(p_m), \\ 
~~~~~~\{\Phi_{2q-1,2q-2}(p_{m'}) + rank(s_i, p_m) - rank(s_i, p_{m'}) : s_i \in S_m' \wedge f(s_i, p_{m'})=1\}\}$.
\end{minipage}
\end{eqnarray}
Let $\rho'_q(p_m)$ be the profile computed at $p_m$ after the first sub-loop during the $q^{th}$ iteration of the main loop of the \sfp{Get-max-aug} algorithm (i.e. at Line \ref{sec_loop} during the $q^{th}$ iteration). Then
\begin{eqnarray}
\label{lbl_b}
\centering
\begin{minipage}{0.9\linewidth}
$\rho'_{q}(p_m) = \max_{\succ_L}\{\rho_{q-1}(p_m), \\ 
~~~~~~\{\rho_{q-1}(p_{m'}) + rank(s_i, p_m) - rank(s_i, p_{m'}) : s_i \in S_m' \wedge f(s_i, p_{m'})=1\}\}$.
\end{minipage}
\end{eqnarray}
By the induction hypothesis, $\rho_{q-1}(p_m) \succeq_L \Phi_{2q-1,2q-2}(p_m)$. Thus:
\begin{eqnarray*}
\centering
\begin{minipage}{0.9\linewidth}
$\rho'_{q}(p_m) = \max_{\succ_L}\{\rho_{q-1}(p_m), \{\rho_{q-1}(p_{m'}) + rank(s_i, p_m) - rank(s_i, p_{m'}):\\
~~~~~~~~~~~~~~  s_i \in S_m' \wedge f(s_i, p_{m'})=1\}\}$. (by equation \ref{lbl_b}).\\
$\succeq_L \max_{\succ_L}\{\Phi_{2q-1,2q-2}(p_m), \{\Phi_{2q-1,2q-2}(p_{m'}) + rank(s_i, p_m) - rank(s_i, p_{m'}):\\
~~~~~~~~~~~~~~ s_i \in S_m' \wedge f(s_i, p_{m'})=1\}\}$. (by the inductive hypothesis)\\
$=\Phi_{2q+1,2q-2}(p_m)$. (by equation \ref{lbl_a})
\end{minipage}
\end{eqnarray*}
Therefore:
\begin{eqnarray}
\label{lbl_c}
\centering
\begin{minipage}{0.9\linewidth}
$\rho'_{q}(p_m) \succeq_L \Phi_{2q+1,2q-2}(p_m)$.
\end{minipage}
\end{eqnarray}
Again, if there does not exist a partial augmenting path from an exposed student to $p_m$, of length $\leq (2q+1, 2q)$ and with a better profile than $\Phi_{2q+1, 2q-2}(p_m)$, then $\Phi_{2q+1,2q}(p_m) = \Phi_{2q+1,2q-2}(p_m)$. 
Otherwise there exists a partial augmenting path from an exposed student to $p_m$ of length $(2q+1, 2q)$ with a better profile than $\Phi_{2q+1,2q-2}(p_m)$. We can therefore  note the following identity involving  $\Phi_{2q+1,2q}(p_m)$:
\begin{eqnarray}
\label{lbl_d}
\centering
\begin{minipage}{0.9\linewidth}
$\Phi_{2q+1,2q}(p_m) = \max_{\succ_L}\{\Phi_{2q+1,2q-2}(p_m), \\ 
~~~~~~\{\Phi_{2q+1,2q-2}(p_{m'}): l_k = l(p_m) \wedge p_{m'} \in P_k' \wedge f(p_m, l_k)>0 \wedge f(l_k, v_t)=d_k^+\}\}$.
\end{minipage} 
\end{eqnarray}
After the $q^{th}$ iteration of the main loop has completed, we have:
\begin{eqnarray}
\label{lbl_e}
\centering
\begin{minipage}{0.9\linewidth}
$\rho_q(p_m) = \max_{\succ_L}\{\rho'_q(p_m), \\ 
~~~~~~~~~~~~~~\{\rho'_q(p_{m'}): l_k = l(p_m) \wedge p_{m'} \in P_k' \wedge f(p_m, l_k)>0 \wedge f(l_k, v_t)=d_k^+\}\}$.
\end{minipage} 
\end{eqnarray}
We observe that the extra condition ($pred(p_m) \neq l_k$) in Line \ref{extra_condition} of the second loop, does not affect the correctness of equation \ref{lbl_e}. Suppose $pred(p_m) = l_k$, then $\rho(p_m)$ must have been updated during the $q^{th}$ iteration of the second loop (or during a previous iteration and has remained unchanged) by some project profile $\rho(p_j')$. 
Thus setting $\rho(p_j) = \rho(p_m)$ and $pred(l_k) = p_m$ would be incorrect as $p_j'$ is now the source of $\rho(p_m)$ and not $p_m$. Moreover if indeed $\rho(p_m) = \rho(p_j') \succ_L \rho(p_j)$ then $p_j'$ would be encountered later on during the iteration of the second loop.
\begin{eqnarray*}
\centering
\begin{minipage}{0.9\linewidth}
$\rho_q(p_m) = \max_{\succ_L}\{\rho'_q(p_m), \\ 
~~~~~~~~~~~~~~\{\rho'_q(p_{m'}): l_k = l(p_m) \wedge p_{m'} \in P_k' \wedge f(p_m, l_k)>0 \wedge f(l_k, v_t)=d_k^+\}\}$.\\
$\succeq_L  \max_{\succ_L}\{\Phi_{2q+1,2q-2}(p_m), \{\Phi_{2q+1,2q-2}(p_{m'}): l_k = l(p_m) \wedge\\
~~~~~~~~~~~~~~p_{m'} \in P_k' \wedge f(p_m, l_k)>0 \wedge f(l_k, v_t)=d_k^+\}\}$ (from equation \ref{lbl_c})\\
$=\Phi_{2q+1,2q}(p_m)$ (by equation \ref{lbl_d}).
\end{minipage} 
\end{eqnarray*}
Therefore:
\begin{eqnarray*}
\centering
\begin{minipage}{0.9\linewidth}
$\rho_{q}(p_m) \succeq_L \Phi_{2q+1,2q}(p_m)$.
\end{minipage}
\end{eqnarray*}

But any partial augmenting path from an exposed student to $p_j$ with respect to flow $f$ can have length at most $(2k+1,2k)$. Thus $\rho(p_j) = \Phi_{2k+1,2k}(p_j)$ after $k$ iterations of the main loop.

Finally we show that a partial augmenting path $P'$ (and subsequently a full augmenting path) can be constructed by following the $pred$ values of projects and lecturers and the matched edges in $M(f)$ starting from some exposed project $p_j$ with the maximum $\rho(p_j)$ profile, and ending at some exposed student (i.e. we show that such a path is continuous and contains no cycle).

Suppose for a contradiction that such a path $P'$ contained a cycle $C$. 
Then at some step $X$ during the execution of the algorithm, $C$ would have been formed when, for some project $p_j$, either (i) $pred(p_j)$ was set to some student $s_i$ or (ii) $pred(p_j)$ was set to some lecturer $l_k$. Let $P''$ be any path in $N(I)$. We may extend our definitions for the profile of a matching and a partial augmenting path to cover the profile of any path in $N(I)$ as follows:
\begin{eqnarray*}
\centering
\begin{minipage}{0.8\linewidth}
$\rho(P'') = O_R + \sum\{rank(s_i, p_j) : (s_i, p_j) \in P'' \cap E_2 \wedge f(s_i, p_j)=0\} - \\
~~~~~~~~~~~~\sum\{rank(s_i, p_j): (p_j, s_i) \in P'' \cap E_2 \wedge f(s_i, p_j)=1\}$.
\end{minipage}
\end{eqnarray*}

Considering case (i) let $p_m = M(s_i)$.  Also let $\rho'(p_j)$ and $\rho(p_j)$ be the profiles of partial augmenting paths from some exposed student to $p_j$ before and after step $X$ respectively. Then $\rho(p_j) \succ_L \rho'(p_j)$. Also $\rho(p_j) = \rho(p_m) + rank(s_i, p_j) - rank(s_i, p_m)$, i.e.,  $\rho(p_j) = \rho(p_m) + \rho(P'')$ where $P''=\{(s_i, p_j), (s_i, p_m)\}$. Since we can also trace a path through all the other projects in $C$ (using $pred$ values and matched edges) from $p_m$ to $p_j$, it follows that $\rho(p_m) = \rho'(p_j) + \rho(C \backslash \{(s_i, p_j), (s_i, p_m)\})$. Thus $\rho(p_j) = \rho'(p_j) + \rho(C)$. Note that $\rho(C) = \rho(C' \backslash M)  - \rho(C' \cap M)$ and $C' = C \cap E_2$ is the set of edges in $C$ involving only students and projects. As  $\rho(p_j) \succ_L \rho'(p_j)$, it follows that $\rho(C' \backslash M)  \succ_L \rho(C' \cap M)$. But since $|C' \backslash M| = |C' \cap M|$, and lecturer capacities are clearly not violated by the algorithm, a new matching $M' = M \oplus C'$ can be generated such that $\rho(M') \succ_L \rho(M)$ and $|M'| = |M| = |f|$, a contradiction to the fact that $M$ is a greedy $|f|$-matching in $I$.

Considering case (ii) let $p_m = pred(l_k)$. As before let $\rho'(p_j)$ and $\rho(p_j)$ be the profiles of partial augmenting paths from some exposed student to $p_j$ before and after step $X$ respectively. Then $\rho(p_j) \succ_L \rho'(p_j)$. Also $\rho(p_j) = \rho(p_m)$. Since we can also trace a path through all the other projects in $C$ (using $pred$ values and matched edges) from $p_m$ to $p_j$, it follows that $\rho(p_m) = \rho'(p_j) + \rho(C \backslash \{(p_j, l_k), (p_m, l_k)\}) = \rho'(p_j) + \rho(C)$. Thus $\rho(p_j) = \rho'(p_j) + \rho(C)$. Note that $\rho(C) = \rho(C' \backslash M)  - \rho(C' \cap M)$ and $C' = C \cap E_2$ is the set of edges in $C$ involving only students and projects. As  $\rho(p_j) \succ_L \rho'(p_j)$, it follows that $\rho(C' \backslash M)  \succ_L \rho(C' \cap M)$. A similar argument to the one presented above shows a contradiction to the fact that $M$ is a greedy $|f|$-matching in $I$.
\end{proof}

From Lemmas \ref{kwa1}, \ref{kwa2} and \ref{kwa3}, we can conclude that the algorithm \sfp{Greedy-max-spa} finds a greedy maximum matching given a {\sc spa} instance. Concerning the complexity of the algorithm, the main loop calls \sfp{Get-max-aug} $\eta$ times where $\eta$ is the size of a maximum cardinality matching in $I$. The first phase of \sfp{Get-max-aug} performs $O(m_2)$ profile comparison operations and $O(n_3)$ initialisation steps for the lecturer $pred$ values where $m_2 = |E_2|$, $n_3 = |\mathcal{L}|$, and each profile comparison step requires $O(R)$ time.  The loop in the main phase of \sfp{Get-max-aug} runs $k$ times where $k$ is the value of the flow obtained at that time. The first and second loops perform $O(m_2)$ and $O(n_2)$ relaxation steps respectively where $n_2 = |\mathcal{P}|$ and each relaxation step requires $O(R)$ time to compare profiles. The final phase of the algorithm performs $O(n_2)$ profile comparisons, each also taking $O(R)$ time. Thus the overall time complexity of the \sfp{Get-max-aug} method is $O(m_2R + n_3 + kR(m_2 + n_2) + n_2R) = O(kR(m_2))$. Thus the overall time complexity of the \sfp{Greedy-max-spa} algorithm is $O(n_1^2Rm_2)$. 

When considering the additional factor of $O(R)$ due to arithmetic on edge weights of $O(n_1^R)$ size, Orlin's algorithm runs in $O(Rm_2^2\log (n_1+n_2) + Rm_2(n_1+n_2)\log^2(n_1+n_2))$ time. Suppose $n_1 \geq n_2$. Then Orlin's algorithm runs in $O(Rm_2^2\log n_1 + n_1Rm_2\log^2n_1)$ time.
If the first term of  Orlin's runtime is larger than the second then our algorithm is slower by a factor of $\frac{n_1^2}{m_2\log n_1} \leq \frac{n_1}{\log n_1}$ as $m_2 \geq n_1$.
If the second term of  Orlin's runtime is larger than the first then our algorithm is slower by a factor of $\frac{n_1}{\log^2 n_1} \leq \frac{n_1}{\log n_1}$.

Now suppose $n_2 > n_1$. Then Orlin's algorithm runs in $O(Rm_2^2\log n_2 + n_2Rm_2\log^2n_2)$ time.
If the first term of  Orlin's runtime is larger than the second then our algorithm is slower by a factor of $\frac{n_1^2}{m_2\log n_2} \leq \frac{n_1}{\log n_2} \leq \frac{n_1}{\log n_1}$ as $m_2 \geq n_1$ and $n_2 > n_1$.
If the second term of  Orlin's runtime is larger than the first then our algorithm is slower by a factor of $\frac{n_1^2}{n_2\log^2 n_2} \leq \frac{n_1}{\log^2 n_2} \leq \frac{n_1}{\log n_1}$ as $n_2 > n_1$.

So our algorithm is slower than  Orlin's by a factor of $\frac{n_1}{\log n_1}$ in all cases.
A straightforward refinement of our algorithm can be made by observing that if no profile is updated during an iteration of the main loop, then no further profile improvements can be made and we can terminate the main loop at this point. We conclude with the following theorem.

\begin{theorem1}
\label{lec}
Given a {\sc spa} instance $I$, a greedy maximum matching in $I$ can be obtained in $O(n_1^2Rm_2)$ time.
\end{theorem1}

%--------------------------------------------------------------------------------------------------------------------------------------------------
\section{Generous maximum matchings in {\sc spa}}
\label{generous}
Analogous to the case for greedy maximum matchings, generous maximum matchings can also be found by modelling  {\sc spa} as a network flow problem. Given a {\sc spa} instance $I$ we define the following terms relating to partial augmenting paths in $N(I)$. For each project $p_j \in \mathcal{P}$, we define the \emph{minimum profile of a partial augmenting path} from $v_s$ through an exposed student to $p_j$ with respect to $\prec_R$, denoted $\Phi'(p_j)$, as follows:
\begin{eqnarray*}
\begin{minipage}{400px}
\centering
$\Phi'(p_j) = \min_{\prec_R}\{\rho(P'): P' $ is a partial augmenting path from $v_s$ to $p_j\}$. 
\end{minipage}
\end{eqnarray*}
If a partial augmenting path $P'$ ending at project $p_j$ can be extended to an augmenting path $P$ by adding edges $(p_j, l(p_j))$ and $(l(p_j), v_t)$ then such an augmenting path is called a \emph{minimum profile augmenting path} if $\rho(P) = \min_{\prec_R}\{\Phi'(p_j) : p_j \in \mathcal{P}\}$. 
A similar approach to that used to find a greedy maximum matching can be adopted in order to find a generous maximum matching. The main \sfp{Greedy-max-spa} algorithm will remain unchanged (we will call it \sfp{Generous-max-spa} for convenience) as the intuition remains to successively find larger generous $k$-matchings until a generous maximum matching is obtained. We however make slight changes to the \sfp{Get-max-aug} algorithm in order to find a minimum profile augmenting path in the network should one exist (the resulting algorithm is then known as \sfp{Get-min-aug}). The changes are as follows.  (i) We replace all occurrences of left domination $\succ_L$ with right domination $\prec_R$. (ii) We also replace all occurrences of negative infinity profile $B_R^-$ with a positive infinity profile $B_R^+$.
%In line \ref{set_neg_profile} we replace the negative infinity profile $B_R^-$ with a positive infinity profile $B_R^+$ when initialising $\rho(p_j)$ for every $p_j \in \mathcal{P}$. 
 (iii) Finally we replace both $\max$ functions (in lines \ref{first_max} and \ref{second_max}) with the $\min$ function. Analogous statements and proofs of Lemmas \ref{kwa1}, \ref{kwa2} and \ref{kwa3} exist in this context. Thus we may conclude with the following theorem concerning the  \sfp{Generous-max-spa} algorithm.

\begin{theorem1}
\label{lec2}
Given a {\sc spa} instance $I$, a generous maximum matching in $I$ can be obtained in $O(n_1^2Rm_2)$ time.
\end{theorem1}

%===============================================================================%
\section{Lecturer lower quotas}
\label{lower-q}
In \scp{spa} problems it is often required that the workload of supervising student projects is evenly spread across the lecturing staff (i.e., that project allocations are \emph{load-balanced} with respect to lecturers). This is important because any project allocation should be seen by lecturers to be fair. Moreover a lecturer's workload may have an effect on her performance in other academic and administrative duties.
One way of achieving some notion of load-balancing with respect to lecturers is to introduce lower quotas. A lower quota on lecturer $l_k$ is the minimum number of students that must be assigned to $l_k$ in any feasible solution. 
We call this extension the \emph{Student/Project Allocation problem with Lecturer lower quotas \scp{(spa-l)}}. In an instance $I$ of \scp{spa-l}, each lecturer $l_k$ has an upper quota $d_k(I)^+ = d_k^+$ and now additionally has a lower quota $d_k^-(I)$ (it will be helpful to indicate specific instances to which these lower quotas refer within the notation). We assume that $d_k^-(I) \geq 0$ and $d_k^+(I) \geq \max\{d_k^-(I), 1\}$. In the \scp{spa-l} context, our definition of a matching as presented in Section \ref{spa_defs} needs to be tightened slightly. A \emph{constrained matching} is a matching $M$ in the \scp{spa} context with the additional property that, for each lecturer $l_k$, $|M(l_k)| \geq d_k^-(I)$. 
A \emph{constrained maximum matching} is a maximum matching taken over the set of constrained matchings in $I$.
Suppose that $L$ is the sum of the lecturer lower quotas in $I$ (i.e. $L = \sum_{l_k \in \mathcal{L}}d_k^-(I)$) and $\eta$ is the size of a maximum matching in $I$\footnote{We will prove that $\eta$ is equal to the size of a maximum constrained matching in Proposition \ref{eta_constrained}}. 
For some $k$ in $(L \leq k \leq \eta$), let $\mathcal{M}'_k$ denote the set of constrained matchings of size $k$ in $I$.  A matching $M\in \mathcal M'_k$ is a \emph{constrained greedy $k$-matching}  if $M$ has lexicographically maximum profile, taken over all matchings in $\mathcal M'_k$. An analogous definition for a \emph{constrained generous $k$-matching} can be made.

Due to the introduction of these lecturer lower quotas, instances of \scp{spa-l} are not guaranteed to admit a feasible solution. 
Thus given an instance $I$ of \scp{spa-l}, we seek to find a constrained greedy or a constrained generous maximum matching should one exist. 
We therefore present results analogous to Lemmas \ref{kwa1}, \ref{kwa2} and \ref{kwa3}. Firstly however, we make the following observations.

\begin{proposition1}
\label{eta_constrained}
Given an \scp{spa-l} instance $I$, the size of a constrained maximum matching (should one exist) in $I$ is equal to the size of a maximum matching in the underlying \scp{spa} instance in $I$.
\end{proposition1}
\begin{proof}
Assume $I$ admits a constrained matching. Then, by dropping the upper quota of each lecturer $l_q \in \mathcal{L}$ from $d_q^+(I)$ to $d_q^-(I)$, and finding a saturating flow in the network obtained from the  resulting instance, we can obtain a matching $M_k$ of size $k$ where $k = \sum_{l_q \in \mathcal{L}}d_q^-$. By returning the lecturer upper quotas to their original values and then successively finding and satisfying standard augmenting paths (starting from $f(M_k)$) we are bound to obtain a constrained maximum matching as lecturers do not lose any assigned students in the process. The absence  of an augmenting path relative to the final flow is proof that the flow (and resulting constrained matching) is maximum. 
\end{proof}

\begin{figure}[t]
\small
\centering
\begin{minipage}[b]{0.6\linewidth}
\centering
\begin{align*}
&\mbox{students' preferences:} ~~~~~~~~~&&\mbox{lecturers' offerings:}\\
&s_1: p_1~~~p_2  &&l_1: \{p_1\} \\
&s_2: p_3~~~p_2 &&l_2: \{p_2\} \\
&s_3: p_3 &&l_3: \{p_3\}
\end{align*}
$c_1= c_3=1$, $d_1^+ = d_3^+ = 1$ and $c_2=d_2^+=2$\\
$d_1^- = d_3^- = 0$ and $d_2^- = 2$
\end{minipage}
\caption{A \scp{spa-l} instance $I$}
\label{spal_instance}
\vspace{-10pt}
\end{figure}

We also observe that a constrained greedy $k$-matching $M_k$ in $I$ need not be a greedy $k$-matching in $I$. That is, there may exist a matching $M_{k}'$ of size $k$ in $I$ such that $M_{k}'$ violates some of its lecturer lower quotas (i.e. $M_{k}'$ is not a constrained matching) and $\rho(M_{k}') \succ_L \rho(M_{k})$. 
Figure \ref{spal_instance} shows a \scp{spa-l} instance whose unique constrained greedy maximum matching is $M = \{(s_1, p_2), (s_2, p_2), (s_3, p_3)\}$ and a greedy maximum matching $M'=\{(s_1, p_1), (s_2, p_2), (s_3, p_3)\}$ such that $\rho(M') \succ_L \rho(M)$.
However it is sufficient to show that, starting from $M_k$, we can successively identify and augment (w.r.t. the incumbent flow) maximum profile augmenting paths in $N(I)$ until a constrained greedy maximum matching is found.  
Next we show that such augmenting paths exist.

\begin{lemma1}
Let $I$ be an instance of {\sc spa-l} and let $\eta$ denote the size of a constrained maximum matching in $I$. Let $k~(1 \leq k < \eta)$ be given and suppose that $M_k$ is a constrained greedy $k$-matching in $I$. Let $N =N(I)$ and $f = f(M_k)$.  Then there exists an augmenting path $P$ with respect to $f$ in $N$ such that if $f'$ is the result of augmenting $f$ along $P$ then $M_{k+1} = M(f')$ is a constrained greedy $(k+1)$-matching in $I$.
\label{greedy-l1}
\end{lemma1}
\begin{proof}
The proof is analogous to that presented for Lemma \ref{kwa1}. We show that considering constrained matchings does not affect most of the arguments presented in  the proof of Lemma \ref{kwa1}. We will deal with the cases where considering constrained matchings may affect the arguments presented in the proof of Lemma \ref{kwa1}.  
Firstly we observe that after cloning the projects in $I$ to form a \scp{spa-l} instance $I' = C(I)$, the process of converting matchings in $I$ to $I'$ and vice versa is unaffected when the matchings considered are constrained. Thus since $M_k$ is a constrained greedy $k$-matching in $I$, $C(M_k)$ is a constrained greedy $k$-matching in $I'$.

Let $M_{k+1}'$ be a constrained greedy $(k+1)$-matching in $I$ (this exists because $k < \eta$). Then $C(M_{k+1}')$ is a constrained greedy $(k+1)$-matching in $I'$. Let $X = C(M_k) \oplus C(M_{k+1}')$. Then each connected component of $X$ is either (i) an alternating cycle, (ii)(a) an even-length alternating path whose end vertices are students, (ii)(b) an even-length alternating path whose end vertices are projects, (iii)(a) an odd-length alternating path whose end edges are in $C(M_k)$ or (iii)(b) an odd-length alternating path whose end edges are in $C(M_{k+1}')$. We firstly show that the procedures used to ``join'' and ``eliminate'' these connected components in Lemma \ref{kwa1} are unaffected when $C(M_k)$ and $C(M_{k+1}')$ are constrained matchings. The even-length components that we firstly consider are:
\begin{enumerate}
\item type (i) and type (ii)(a) alternating paths.
\item compound type (ii)(a) paths.
\end{enumerate}
When considering the elimination of these even-length components (or compound paths), the requirement that the upper quotas of the lecturers involved must not be violated still holds even if the matchings considered are constrained. Moreover the number of students assigned to each lecturer never drops when considering the elimination of these even-length components (or compound paths). 

Let $C(M_{k+1}'')$ be the constrained greedy $(k+1)$-matching obtained from augmenting $C(M_{k+1}')$ along all these even-length paths. Then  $X' = C(M_k) \oplus C(M_{k+1}'')$ consists of a set of compound type(ii)(b) paths, compound type (iii)(a) and compound type (iii)(b) paths. These paths, if considered independently, may lead to some lecturer losing an assigned student when they are used to augment $C(M_k)$ or $C(M_{k+1}'')$. Thus the elimination argument, as presented in the proof of Lemma \ref{kwa1}, does not hold. We modify this argument slightly as follows in the case of constrained matchings. 

We firstly observe that $C(M_k)$ and $C(M_{k+1}'')$ are constrained matchings. Thus augmenting $C(M_k)$ or $C(M_{k+1}'')$ along $X'$ leads to a constrained matching. When all the elements in $X'$ are considered together, no lecturer violates her lower quota. If some lecturer loses a student due to some component of $X'$ and drops below her lower quota, the she must gain an extra student due to another component in $X'$. But since $|C(M_{k+1}'')| = |C(M_{k})| + 1$ there are $q$ compound type (iii)(a) paths and $(q+1)$ compound type (iii)(b) paths in $X'$ for some integer $q$. 
Compound type (ii)(b) components do not affect the size of the matchings. 

 We claim that there exists some compound type (iii)(b) path $P'$ in $X'$ such that when considering all the other components in $X'$ (i.e. $X' \backslash P'$), lecturer upper and lower quotas are not violated and the size of the matchings are unchanged. Thus the elimination arguments presented in the proof of Lemma \ref{kwa1} can be applied to $X' \backslash P'$. 
$P'$ can be extended to end with edge $(l_p, v_t)$ such that $|M_{k+1}''(l_p)| > |M_{k}(l_p)| \geq d_p^-$.
If $C(M_{k+1}''')$ is the constrained greedy $(k+1)$-matching obtained from augmenting $C(M_{k+1}'')$ along $X' \backslash P'$, then $C(M_{k+1}''') \oplus C(M_k) = P'$.  
If such a path $P'$ does not exist then $|M_{k+1}''(l_p)| \leq |M_{k}(l_p)|$ for all $l_p \in \mathcal{L}$, a contradiction. 

The rest of the proof for Lemma \ref{kwa1}, involving the generation of an augmenting path, follows through.
\end{proof}

\begin{lemma1}
Let $f$ be a flow in $N$ and let $M_k = M(f)$. Suppose that $M_k$ is a constrained  greedy $k$-matching. Let $P$ be a maximum profile augmenting path with respect to $f$. Let $f'$ be the flow obtained by augmenting $f$ along $P$. Now let $M_{k+1} = M(f')$. Then $M_{k+1}$ is a constrained greedy $(k+1)$-matching. 
\label{greedy-l2}
\end{lemma1}
\begin{proof}
The proof for Lemma \ref{kwa2} holds  even if $M(f)$ and $M(f')$ are constrained matchings as the number of students assigned to a lecturer never reduces as we augment $f$ along $P$.
\end{proof}

\begin{lemma1}
Given an \scp{spa-l} instance $I$, let $f$ be a flow in $N(I)$ where $k = |f|$ is not the size of a constrained maximum matching in $I$ and $M(f)$ is a constrained greedy $k$-matching in $I$. Algorithm \sfp{Get-max-aug} finds a maximum profile augmenting path in $N(I)$ with respect to $f$.
\label{greedy-l3}
\end{lemma1}
\begin{proof}
We observe that the proof presented for Lemma \ref{kwa3} also holds in this case even if $M(f)$ is a constrained greedy $k$-matching.

The first part of the proof shows that after $q$ iterations of the main loop of \sfp{Get-max-aug} where $0 \leq q \leq k$, $\rho(p_j) \succeq_L \Phi_{2q+1,2q}(p_j)$ for every project $p_j \in \mathcal{P}$ where $\Phi_{2q+1,2q}(p_j)$ is the maximum profile of any partial augmenting path of length $\leq (2q+1,2q)$ from an exposed student to $p_j$. By inspection, we observe that this argument remains unchanged even if $M(f)$ is a constrained matching in $I$.

The second part of the proof shows that  a partial augmenting path $P'$ (and subsequently a full augmenting path) can be constructed by following the $pred$ values of projects and lecturers and the matched edges in $M(f)$ starting from some exposed project $p_j$ with the maximum $\rho(p_j)$ profile going through some exposed student and ending at the source $v_s$.  That is, we show that such a path is continuous and contains no cycle. We prove this by demonstrating that, should a cycle $C$ exist, then augmenting $f$ along $C$ would yield a flow of the same size $f'$ such that $M(f') \succ_L M(f)$ which is a contradiction to the fact that $M(f)$ is a greedy $k-$matching. This result also holds in the case where $M(f)$ is a constrained matching as any cycle found will not cause a lecturer to lose any assigned students and so the above arguments can still be made. 
\end{proof}

\begin{algorithm}[t]
\small
\caption{\sfp{Greedy-max-spa-l}}
\label{gen-spa-l}
\begin{algorithmic}[1]
	\REQUIRE \scp{spa-l} instance $I$;
	\ENSURE return a matching $M$ if one exists or {\bf null} otherwise;
	\STATE copy $I$ to from new instance $I'$;
	\FORALL{lecturer $l_k \in I'$}
		\STATE set $d_k^+(I') = d_k^-(I)$;
		\STATE set $d_k^-(I') = 0$;
	\ENDFOR
	\STATE  \{$I'$ becomes a \scp{spa} instance\}
	\STATE $M' = $ \sfp{Greedy-max-spa}($I'$); \label{saturated_flow} 
	\IF{$|M'| ~=~ \sum_{l_k \in \mathcal{L}}d_k^-(I)$}
		\STATE copy $f(M')$ in $N(I')$ into $f$ in $N(I)$;
		\LOOP
			\STATE $P = $ \sfp{Get-max-aug}$(N(I), f)$;  \label{saturated_flow2}
			\IF {$P \neq null$}
				\STATE augment $f$ along $P$;
			\ELSE
				\STATE return $M(f)$;
			\ENDIF
		\ENDLOOP
	\ELSE
		\STATE return {\bf null};
	\ENDIF
\end{algorithmic}
\normalsize
\end{algorithm}

Given Lemmas \ref{greedy-l1}, \ref{greedy-l2} and \ref{greedy-l3}, the \sfp{Greedy-max-spa} algorithm can be employed as part of an algorithm to find a constrained greedy maximum matching in a \scp{spa-l} instance should one exist. This new algorithm (which we call \sfp{Greedy-max-spa-l}) is presented in Algorithm \ref{gen-spa-l}. The algorithm takes an \scp{spa-l} instance $I$ as input and returns a constrained greedy maximum matching $M$, should one exist, or {\bf null} otherwise.
A \scp{spa} instance $I'$ is constructed from $I$ by setting $d_k^-(I') = 0$ and $d_k^+(I') = d_k^-(I)$ for each lecturer $l_k$. Next we find a greedy maximum matching $M'$ in $I'$ using the \sfp{Greedy-max-spa} algorithm. If $f' = f(M')$ is not a saturating flow (i.e., one in which all edges $(l_k, v_t) \in E_4$ are saturated), then $I$ admits no constrained matching and we return {\bf null}. Otherwise we augment flow $f$ in $N(I)$ by calling the \sfp{Get-max-aug} algorithm, where $f$ is the flow in $N(I)$ obtained from cloning $f'$ in $N(I')$. We continuously augment the flow until no augmenting path exists. The matching $M = M(f)$ obtained from the resulting flow $f$ is a greedy maximum constrained matching in $I$. Constrained generous maximum matchings can also be found in a similar way.
We conclude with the following theorem.

\begin{theorem1}
Given a \scp{spa-l} instance $I$, a constrained greedy maximum matching and a constrained generous maximum matching in $I$ can be obtained, should one exist, in $O(n_1^2Rm_2)$ time.
\end{theorem1}
\begin{proof}
Firstly we show that the matching $M'$ obtained in Line \ref{saturated_flow} of the \sfp{Greedy-max-spa-l} algorithm is a constrained greedy $|f|$-matching in $I$. Suppose otherwise and some other constrained matching $M''$ of the same size exists in $I$ such that $\rho(M'') \succ_L \rho(M')$. Then since $|f| =  \sum_{l_k \in \mathcal{L}}d_k^-(I)$, every lecturer has exactly the same number of assigned students in $M'$ and $M''$ so $M''$ is a valid matching in $I'$. This contradicts the fact that $M'$ is a greedy maximum matching in $I'$. 

Lemmas \ref{greedy-l1}, \ref{greedy-l2} and \ref{greedy-l3} prove that once we obtain a constrained greedy $|f|$-matching in $I$ (should one exist), the rest of the algorithm finds a maximum constrained greedy maximum matching in $I$. 

For finding a constrained generous maximum matching we simply replace the call to \sfp{Greedy\-max\-spa} in Line \ref{saturated_flow} and the call to \sfp{Get-max-aug} in Line \ref{saturated_flow2} of the \sfp{Greedy-max-spa-l} algorithm with a call to the  \sfp{Generous-max-spa} and the  \sfp{Get-min-aug} algorithms respectively as described in Section \ref{generous}. 
\end{proof}

%--------------------------------------------------------------------------------------------------------------------------------------------------
\section{Empirical evaluation}
\label{spa_impl}
\subsection{Introduction}
\label{spa_impl_intro}
The \sfp{Greedy-Max-Spa} and \sfp{Generous-Max-Spa} algorithms were implemented in Java and evaluated empirically. 
%The \scp{spa} IP models were also encoded and solved using the IBM CPLEX 12.5.1 IP solver. 
In this section, we present results from empirical evaluations carried out on the algorithm implementations using both real-world and randomly-generated data.
Results from the implemented algorithms were compared with those produced by an IP model of \scp{spa} in order to improve our confidence in the correctness of both implementations.
We also investigate the feasibility issues that will be faced if a Min-Cost-Max-Flow (\scp{mcmf}) approach (as suggested in \cite{MU14}) is to be used when solving instances of \scp{spa} involving large numbers of students and projects or were students have long preference lists.
Other experiments carried out involve varying certain properties of the randomly-generated \scp{spa} instances while measuring the runtime of the algorithms and the size, degree and cost of the matchings produced. 

An instance generator was used to construct random {\sc spa} instances which served as input for the algorithm implementations. This generator can be configured to vary certain properties of the {\sc spa} instances produced as follows:

\begin{enumerate}
\item The number of students $n_1$ (with a default value of $n_1 = 100$). The number of projects and lecturers are set to $n_2 = 0.3n_1$ and $n_3 = 0.3n_1$ respectively.
\item The minimum $R_{min}$ and maximum $R_{max}$ length of any student's preference list (with default values $R_{min} = R_{max} = 10$).
\item The popularity $\lambda$ of the projects, as measured by the ratio between the number of students applying for one of the most popular projects and the number of students applying for one of the least popular projects (default value of $5$).
\item The total capacity of the projects $C_{\mathcal{P}}$ and lecturers $C_{\mathcal{L}}$. These capacities were not divided evenly amongst the projects and lecturers involved (default values are $C_{\mathcal{P}} = 1.2n_1$ and $C_{\mathcal{L}} = 1.2n_1$).
\item The tie density $t_d~(0 \leq t_d \leq 1)$ of the students' preference list. This is the probability that some project is tied with the one preceding it  on some student's preference list (default value is $t_d=0$).
\item The total project and lecturer lower quotas $L_{\mathcal{P}}$ and $L_{\mathcal{L}}$ respectively.  These lower quotas were divided evenly amongst the projects and lecturers involved (default values are $L_{\mathcal{P}} = L_{\mathcal{L}} = 0$).
\end{enumerate}

We also created {\sc spa} instances from anonymised data obtained from previous runs of the student-project allocation scheme at the School of Computing Science, University of Glasgow and solved them using the implemented algorithms. We measured the runtime taken by the algorithms as well as the size, cost and degree of the matchings obtained.
Experiments were carried out on a Windows machine with 4 Intel(R) Core(R) i5-2400 CPUs at 3.1GHz and 8GB RAM. 

In the following subsections we present results obtained from the empirical evaluations carried out. In Section \ref{spa1_correctness} we present the results of correctness tests carried out by comparing results obtained from IP models of \scp{spa} and implemented algorithms. In Section \ref{spa_feas} we demonstrate when the \scp{mcmf} approach becomes infeasible in practice.  In Section \ref{real_data} we present results from running the algorithms against real-world \scp{spa} instances. In Section \ref{spa_random} we vary certain properties of randomly generated \scp{spa} instances while measuring the runtime of the algorithms and the size, degree and cost of the matchings produced. We make some concluding remarks in Section \ref{emp_conclusions}.

%--------------------------------------------------------------------------------------------------------------------------------------------------
\subsection{Testing for correctness}
\label{spa1_correctness}
Although the  \sfp{Greedy-Max-Spa} and \sfp{Generous-Max-Spa} algorithms have been proven to be correct (See Theorems \ref{lec} and \ref{lec2}), bugs may still exist in the implementations. In order to improve our confidence in any empirical results obtained as part of an experimental evaluation of the algorithms' performance, we compared results from the implemented algorithms with those obtained from IP models of \scp{spa}. For each value of $n_1$ in the range $n_1 \in \{20, 40, 60, ..., 200, 300, 400, ..., 1000\}$, $10,000$ random {\sc spa} instances were generated and solved using both methods.
For each {\sc spa} instance generated, $R_{min} = R_{max} = 10$ (henceforth we refer to $R_{min} = R_{max}$ as $R$). The profiles of the resulting matchings were then compared and observed to be identical for all the instances generated. The resulting matchings were also tested to ensure they obeyed all the upper quota constraints for lecturers and projects. 
These correctness tests show that our implementations are likely to be correct.

%--------------------------------------------------------------------------------------------------------------------------------------------------
\subsection{Feasibility analysis of the \scp{mcmf} approach}
\label{spa_feas}
We implemented an algorithm for finding a minimum cost maximum flow in a given network. As stated in \cite{Abr03,MU14}, by the appropriate assignment of edge costs/weights in the underlying network $N(I)$ of a \scp{spa} instance $I$, a minimum cost maximum flow algorithm can be used to find greedy and generous maximum matchings in $I$. We argued  that this approach (as described in \cite{Abr03,MU14}) would be infeasible due to the floating-point inaccuracies caused by the assignment of exponentially large edge costs/weights in the network. In this section we investigate this claim experimentally and demonstrate the feasibility issues that arise when using various Java data types  to represent these edge weights. 

Firstly we describe the cost functions required by a minimum cost maximum flow algorithm to find greedy and generous maximum matchings.  For finding greedy maximum matchings we set the cost of an edge between a student $s_i$ and a project $p_j$ as $n_1^{R-1}-n_1^{R-k}$ where $k=rank(s_i, p_j)$. For finding generous maximum matchings  we set the cost of an edge between a student $s_i$ and a project $p_j$ as $n_1^{k-1}$ where $k=rank(s_i, p_j)$. The cost for all other edges in the network are set to $0$. 

For the \scp{mcmf} approach, we define an instance as infeasible if the matching  produced is not optimal with respect to the greedy or generous criteria (when compared with optimal results produced by the \sfp{Greedy-Max-Spa} and \sfp{Generous-Max-Spa} algorithms and  CPLEX). We also consider an instance infeasible if the JVM runs out of memory when using the \scp{mcmf} algorithm but does not when using the \sfp{Greedy-Max-Spa} and \sfp{Generous-Max-Spa} algorithms. 

\begin{figure}[!h]
\centering
\small
\begin{minipage}[b]{0.75\linewidth}
\centering
\setlength\fboxsep{0pt}
\setlength\fboxrule{0pt}
\fbox{\includegraphics[width=200pt]{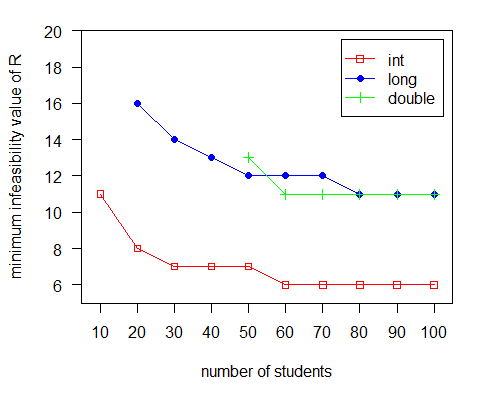}}
\caption{\scp{mcmf} feasibility results}
\label{mcmf-feasibility}
\end{minipage}
\end{figure}

Figure \ref{mcmf-feasibility} shows the feasibility results using three Java data types. For each value of $n_1$ (number of students) in the range $n_1 \in \{10, 20, 30, ..., 100\}$ and for each value of $R$ (length of each student's preference list) in the range $R \in \{5, 6, ..., \min\{20, 1.2n_1\}\}$, we generated $1000$ random \scp{spa} instances and solved them using the \scp{mcmf} approach and the \sfp{Greedy-Max-Spa} algorithm. The graph shows the value of $R$ at which infeasible solutions were first encountered. As expected, this number drops as we increase the instance size. Due to their greater precision that the \sfp{long} and \sfp{double} data types (when compared with \sfp{int}), we see that they handle much larger instances before encountering infeasibility issues. 
All instances tested for $n_1 = 10$ when using \sfp{long} and $n_1 \in \{10, 20, 30, 40\}$ when using \sfp{double} produced optimal matchings. This is probably because we do not yet encounter range errors (in the case of \sfp{long}) and precision errors (in the case of \sfp{double}) when solving these instances.
The relatively low values of $R$ and $n_1$ observed where infeasibility prevails (e.g., $n_1=60, R=6$ for the \sfp{int} type) reinforces our argument that approaches based on \scp{mcmf} which employ these exponentially large edge weights are not scalable. 
 
%--------------------------------------------------------------------------------------------------------------------------------------------------
\subsection{Real-world data}
\label{real_data}
{\sc spa} instances derived from anonymised data obtained from previous runs of the student-project allocation scheme at the School of Computing Science, University of Glasgow were created and solved using the \sfp{Greedy-max-spa} algorithm. This section discusses some of the results obtained.  Table \ref{dcs-spa-ins} shows the properties of the generated {\sc spa} instances (with lecturer capacities not being considered in the $07/08$ and $08/09$ sessions) and Table \ref{dcs-spa-res} shows details of various profile-based optimal matchings found.

\begin{table}[!h]
\centering
\begin{tabular}{|l|llllll|}
\hline
Session &$n_1$ &$n_2$ &$n_3$ &$R$ &$C_{\mathcal{P}}$ &$C_{\mathcal{L}}$ \\ \hline \hline
14/15 &51 &147 &37 &6 &147 & 80\\  \hline
13/14 &51 &155 &40 &5 &155 & 77\\  \hline
12/13 &38 &133 &34 &5 &133 & 63\\  \hline 
11/12 &31 &103 &26 &5 &103 & 62\\  \hline
10/11 &34 &63 &29  &5 &63 & 66\\  \hline
09/10 &32 &102 &28  &5 &102 & 72\\  \hline
08/09$^*$ &37 &56 & - &5 &56 & 56\\  \hline
07/08$^*$ &35 &61 & -  &5 &61 & 61\\  \hline
\end{tabular}
\caption{Real-world {\sc spa} instances}
\label{dcs-spa-ins}
\end{table}

\begin{table}[!h]
\centering
\begin{tabular}{|l|l|ll|ll|ll|}
\hline
Session &$|M|$&Greedy & & Generous & & Min-Cost & \\ \hline
 &&Profile &Cost & Profile &Cost & Profile &Cost \\ \hline \hline
 14/15 &51 & $(30,7,1,5,5,3)$ & $110$ & $(16,16,9,6,4)$ & $119$ &$(28,11,3,5,2,2)$ & $101$ \\  \hline
13/14 &51 & $(26,7,4,6,8)$ & $116$ & $(15,18,9,6,3)$ & $117$ &$(23,12,5,6,5)$ & $111$ \\  \hline
12/13 &38 & $(26,6,3,2,1)$ & $60$ & $(21,13,4)$ & $59$ &$(23,11,3,1)$ & $58$ \\  \hline 
11/12 &31 & $(22,6,2,1)$ & $44$ & $(20,9,2)$ & $44$ &$(20,9,2)$ & $44$ \\  \hline
10/11 &34 & $(25,4,3,1,1)$ & $51$ & $(21,9,4)$ & $51$ &$(24,5,4,1)$ & $50$ \\  \hline
09/10 &32 & $(23,4,2,2,1)$ & $50$ & $(19,10,3)$ & $48$ &$(20,9,2,1)$ & $48$ \\  \hline
08/09$^*$ &37 & $(26,6,2,1,2)$ & $58$ & $(23,11,3)$ & $54$ &$(23,11,3)$ & $54$ \\  \hline
07/08$^*$ &35 & $(20,9,5,0,1)$ & $58$ & $(17,14,4)$ & $57$ &$(17,14,4)$ & $57$ \\  \hline
\end{tabular}
\caption{Real-world {\sc spa} results}
\label{dcs-spa-res}
\end{table}

The results demonstrate a  drawback in adopting the greedy optimisation criterion, namely that some students may have projects that are far down their preference lists. In all but the $2011/2012$ session, at least one student had her worst-choice project in a greedy maximum matching. In the $2013/2014$ session the number of students with their worst-choice project is reasonably high and so the greedy maximum matching would probably not be selected for that year.

The degree of generous maximum matchings are usually less than the others (obviously they are never greater). This is usually an attractive property in such matching schemes. In all the years considered apart from the $2013/2014$ session all students got their third choice project or better in the generous maximum matchings produced. However in the $2013/2014$ session applying the generous optimality criterion did not improve on the degree of the matchings produced. 

One of the major advantages of the minimum cost maximum matching optimality criterion is that in a certain sense it is more ``egalitarian''. Minimising the overall cost of the matchings produced is also a very natural objective. It may be considered a disadvantage if matchings obtained by adopting the profile-based optimality criteria have significantly larger costs than the minimum obtainable cost. However, from the results obtained on these real-world datasets, there is very little difference between the costs of the greedy and generous maximum matchings and the minimum obtainable costs (except, once again, for the $2013/2014$ session). Thus we can choose one of the profile-based optimal matchings with some confidence that it is ``almost" of minimum cost. In Section \ref{spa_random} we consider these differences on multiple randomly generated {\sc spa} instances.

%--------------------------------------------------------------------------------------------------------------------------------------------------
\subsection{Randomly-generated instances}
\label{spa_random}
\subsubsection{Introduction}
This section discusses some of the results obtained by varying certain properties of the randomly generated {\sc spa} instances and measuring the cost, size and  degree of the matchings produced.  For each instance generated we found a greedy maximum matching, a generous maximum matching and a  minimum cost maximum matching. 

%--------------------------------------------------------------------------------------------------------------------------------------------------
\subsubsection{Varying the number of students}
Keeping $R$ constant, we investigated the effects of increasing the number of students $n_1$ (and by implication $n_2$, $n_3$, $C_{\mathcal{P}}$ and $C_{\mathcal{L}}$ using the default dependencies listed in Section \ref{spa_impl_intro}) on the degree, cost and size of the matchings produced as well as the time taken to find these matchings. 
For each value of $n_1$ in the range $n_1 \in \{100, 200, 300, ..., 700\}$ we generated and solved $100$ random \scp{spa} instances. 

\begin{figure}[!h]
\centering
\small
\begin{minipage}[b]{0.45\linewidth}
\centering
\setlength\fboxsep{0pt}
\setlength\fboxrule{0pt}
\fbox{\includegraphics[width=180pt]{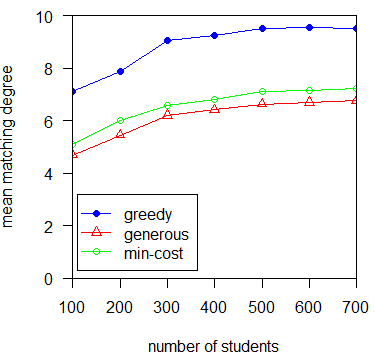}}
\caption{Mean matching degree vs $n_1$}
\label{vary_size_4}
\end{minipage}
\hspace{1cm}
\begin{minipage}[b]{0.45\linewidth}
\centering
\setlength\fboxsep{0pt}
\setlength\fboxrule{0pt}
\fbox{\includegraphics[width=180pt]{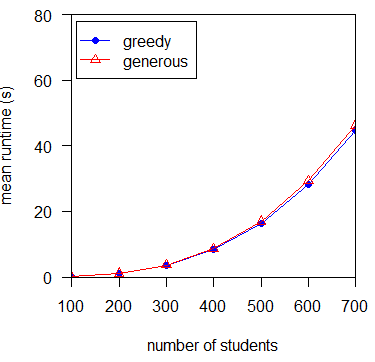}}
\caption{Mean algorithm runtime vs $n_1$}
\label{vary_size_5}
\end{minipage}
\end{figure}

Figure \ref{vary_size_4}  shows the way the mean degree varies as we increase the number of students. The mean degrees of the greedy maximum   matchings are the highest of the three with mean values $\geq 8$ for $n_1>200$. As expected generous maximum matchings have the smallest degree, which rises slowly from about $4.8$ to $6.5$. An interesting observation is that the mean degree does not steeply rise as we increase the number of students. Also the mean degree for the minimum cost maximum matching is closer to the generous maximum matching degree than that of the greedy maximum matching. This is probably due to the fact that the cost function ($rank$ in this case) is greater for higher degrees than lower ones, so, in some way, by minimising the cost, we are also seeking matchings with fewer students matched to projects that are father down their preference lists (i.e. have higher ranks). 

Figure \ref{vary_size_5} shows how long it takes to find both profile-based optimal matchings. 
The main observation is that both \sfp{Greedy-max-spa} and \sfp{Generous-max-spa} algorithms are scalable and can handle decent-sized instances in reasonable times. 
%As in the case where preference list lengths were varied, we observe that it took longer to find a greedy maximum matching than a generous maximum matching. It is also not surprising that the minimum cost matching took even less (as it does not involve these hierarchical objective functions). 

\begin{figure}[!h]
\centering
\small
\begin{minipage}[b]{0.45\linewidth}
\centering
\setlength\fboxsep{0pt}
\setlength\fboxrule{0pt}
\fbox{\includegraphics[width=200pt]{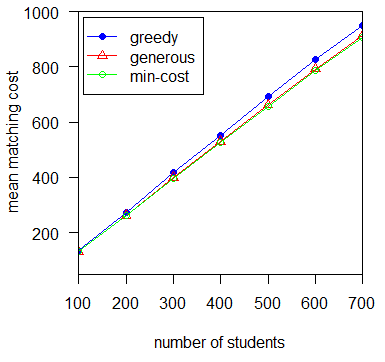}}
\caption{Mean matching cost vs $n_1$}
\label{vary_size_6}
\end{minipage}
\end{figure}

Figure \ref{vary_size_6} shows how the cost of the matchings generated vary with the number of students. The cost seems to grow proportionally with the number of students. We observe that greedy maximum matchings have larger costs than generous and minimum cost maximum matchings. This corresponds to the mean degree curves shown in Figures \ref{vary_size_4} where greedy maximum matchings tend to match some students to projects further down their preference list thus adding to the cost of the matching. The average size of the matchings produced was very close to $n_1$ for all values of $n_1$ tested.

%--------------------------------------------------------------------------------------------------------------------------------------------------
\subsubsection{Varying preference list length}
The length of students' preference list is one property that can be varied easily in practice (in the {\sc spa} context, it is often feasible to ask students to rank more projects if required). So, will increasing the length of the preference lists affect the quality of the matchings produced or the time taken to find them? 
For each value of $R$ in the range $R \in \{1, 2, 3, ..., 10\}$ we tested this by varying the  preference list lengths of $1,000$ randomly generated {\sc spa} instances. Each instance had $n_1=100$ students (with $n_2$, $n_3$, $C_{\mathcal{P}}$ and $C_{\mathcal{L}}$ all assigned their default values).

\begin{figure}[!h]
\centering
\small
\begin{minipage}[b]{0.45\linewidth}
\centering
\setlength\fboxsep{0pt}
\setlength\fboxrule{0pt}
\fbox{\includegraphics[width=180pt]{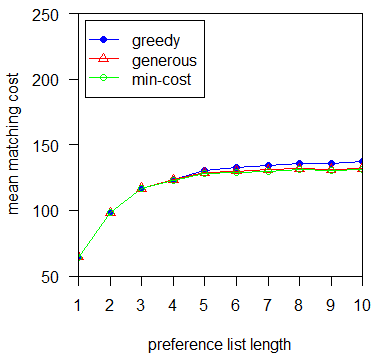}}
\caption{Mean matching cost vs $R$}
\label{vary_pref_1}
\end{minipage}
\hspace{1cm}
\begin{minipage}[b]{0.45\linewidth}
\centering
\setlength\fboxsep{0pt}
\setlength\fboxrule{0pt}
\fbox{\includegraphics[width=180pt]{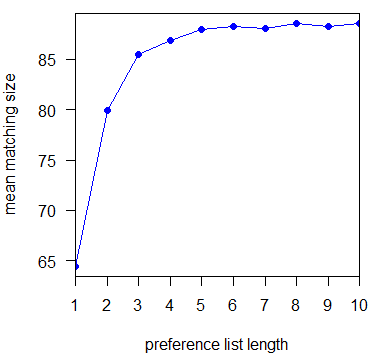}}
\caption{Mean matching size vs $R$}
\label{vary_pref_2}
\end{minipage}
\end{figure}

Figure \ref{vary_pref_1} shows how the mean cost of the matchings obtained varied as we increased the preference list lengths. For the profile-based optimal matchings, the mean cost rises steeply from $R=1$ to $R=4$ but seems to level off beyond that. We observe that the overall cost of the matchings produced does not significantly change for $R>5$.  Thus asking students to submit preference lists greater than $5$ will not significantly affect the overall quality of the generous and minimum cost maximum matchings obtained. Once again we observe a difference between  the cost of the greedy maximum matchings and the other two. 

\begin{figure}[!h]
\centering
\small
\begin{minipage}[b]{0.475\linewidth}
\centering
\setlength\fboxsep{0pt}
\setlength\fboxrule{0pt}
\fbox{\includegraphics[width=200pt]{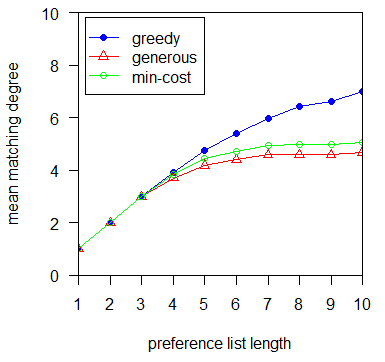}}
\caption{Mean matching degree vs $R$}
\label{vary_pref_3}
\end{minipage}
\end{figure}

Figure \ref{vary_pref_2} also shows an important trend as it highlights the value of $R$ beyond which there is little increase in the mean matching size of profile-based optimal matchings. For the instances generated in this experiment, that value is $R=5$. Thus asking students to submit preference lists of length greater than $5$ will not significantly affect the overall size of maximum matchings obtained. Figure \ref{vary_pref_3} shows how the mean degree of the matchings varied as we increased preference list length. For values of $R \leq 3$ all matchings have the same mean degree as it is likely that some student gets her 3rd choice in each of these matchings.  The curve for minimum cost maximum matchings is closer (with respect to degree) to that of generous maximum matchings (obviously generous maximum matchings have lower degrees in general). They both seem to rise steeply for $R \leq 5$ and then level off at $R=7$ and beyond. Thus asking students to submit preference lists greater than $7$ will not significantly affect the overall degree of generous and minimum cost maximum matchings obtained. As expected, greedy maximum matchings had the highest degrees. For $R > 5$, the mean degree for greedy maximum matchings  does not level off but continues to grow fairly steeply. 

Finally we consider how long it takes for the implemented algorithms to find their solutions. In general, the algorithms all seem to handle \scp{spa} instances with relatively long preference lists ($R=10$) in reasonable time ($< 1.5s$).

%--------------------------------------------------------------------------------------------------------------------------------------------------
\subsubsection{Varying project popularity}
\begin{figure}[!h]
\centering
\small
\begin{minipage}[b]{0.45\linewidth}
\centering
\setlength\fboxsep{0pt}
\setlength\fboxrule{0pt}
\fbox{\includegraphics[width=180pt]{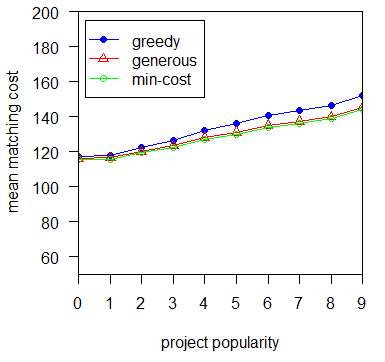}}
\caption{Mean matching cost vs popularity}
\label{vary_pop_1}
\end{minipage}
\hspace{1cm}
\begin{minipage}[b]{0.45\linewidth}
\centering
\setlength\fboxsep{0pt}
\setlength\fboxrule{0pt}
\fbox{\includegraphics[width=180pt]{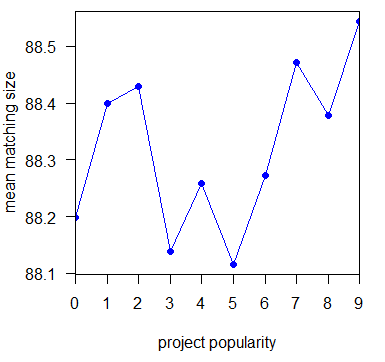}}
\caption{Mean matching size vs popularity}
\label{vary_pop_2}
\end{minipage}
\end{figure}

Not all projects will be equally popular and so it is worth investigating the effects the relative popularity $\lambda$ of the projects may have on the size and quality of the matchings produced. For these experiments, we set $n_1=100$ (with all the other default values) and varied the popularity of the projects involved from $0$ to $9$ in steps of $1$, generating $1,000$ random instances for each popularity value. 
From Figure \ref{vary_pop_1} we see that the cost of the matchings produced gradually increases as we increase the popularity ratio with the cost of the greedy maximum matching being slightly higher than the others (in line with other observations). From Figure \ref{vary_pop_2} we observe no clear trend in the size of the matchings produced as we vary the popularity ratio.

\begin{figure}[!h]
\centering
\small
\begin{minipage}[b]{0.45\linewidth}
\centering
\setlength\fboxsep{0pt}
\setlength\fboxrule{0pt}
\fbox{\includegraphics[width=200pt]{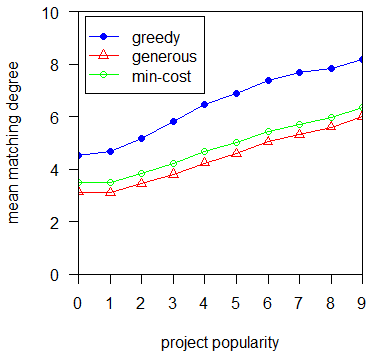}}
\caption{Mean matching degree vs popularity}
\label{vary_pop_3}
\end{minipage}
\end{figure}

Figure \ref{vary_pop_3} shows the gaps between the mean degree of matchings produced using the various algorithms. Once again we see the mean degrees for the  minimum cost and generous maximum matchings being considerably lower than that of the generous maximum matchings as the popularity ratio increases. Runtimes for the \sfp{Greedy-max-spa} and \sfp{Generous-max-spa} algorithms were less than $0.25s$. 

%--------------------------------------------------------------------------------------------------------------------------------------------------
\subsection{Concluding remarks}
\label{emp_conclusions}
Table \ref{tbl_vary_pref_3} gives a breakdown of the profiles of $10,000$ randomly generated {\sc spa} instances of size $n_1=100$ with preference list length $R=10$. It shows the percentage of students with their first choice projects, second choice projects, and so on for greedy, generous and minimum cost maximum matchings (represented by $M_1$, $M_2$ and $M_3$ respectively). Although the choice of which profile-based optimal matching is best will, in practice, be problem-specific, the results (as presented in Sections \ref{real_data} and \ref{spa_random}) give us a general idea of the strengths and weaknesses of the various optimality criteria. We summarise these points below.

\begin{table}[!h]
\centering
\begin{tabular}{|rlllllllllll|}
\hline
 &$1$st &$2$nd &$3$rd &$4$th &$5$th &$6$th &$7$th &$8$th &$9$th &$10$th & Cost \\ \hline \hline
$M_1$: &$67.94$&$18.16$&$7.00$&$3.14$&$1.63$&$0.90$&$0.55$&$0.34$&$0.21$&$0.14$&$161.23$ \\  \hline
$M_2$: &$53.07$&$38.24$&$8.68$&$0.00$&$0.00$&$0.00$&$0.00$&$0.00$&$0.00$&$0.00$&$155.62$ \\  \hline
$M_3$: &$61.88$&$26.84$&$9.31$&$1.85$&$0.13$&$0.00$&$0.00$&$0.00$&$0.00$&$0.00$&$151.51$ \\  \hline
\end{tabular}
\caption{Mean matching profile and cost}
\label{tbl_vary_pref_3}
\end{table}

With greedy maximum matchings we increase the percentage of students that are \emph{happy} with their assigned projects (i.e., obtain their first choice).  A rough estimate of how much better a greedy maximum matching is compared with other profile-based optimal matchings is the difference in the number of first-choice projects.  Table \ref{tbl_vary_pref_3} shows that the percentage of students with their first-choice project is higher when compared with minimum cost maximum matchings (by $6.06\%$) and significantly higher when compared with greedy maximum matchings (by $14.87\%$). However this is achieved at the risk of also increasing the percentage of students who are \emph{disappointed} with their assigned projects (we say a student $s_i$ is disappointed with $p_j = M(s_i)$ if $rank(s_i, p_j) > \lceil R/3 \rceil$).

With generous maximum matchings we reduce the percentage of disappointed students.  A rough estimate of how much better off a student is in a generous maximum matching  compared with a greedy maximum matching, is the difference in the \emph{degree} of the matchings.  Table \ref{tbl_vary_pref_3} shows a significant improvement in the degree as we move from greedy maximum matchings (with some matchings having a degree of $10$) and generous maximum matchings (with all matchings having a degree $\leq 3$).  Although this is usually a very attractive property, this is achieved without considering the percentage of students who are happy with their assignments. Interestingly the generous criterion will continue to attempt to minimise the number of students matched to their $n$th choice project even as $n$ tends to $1$. This motivates a hybrid version of profile-based optimality where we initially adopt the generous criterion and, at some point (say for $r$th choice projects where $r \leq 3$), switch to the greedy criterion. 

Often the profile of a minimum cost maximum matching lies ``in between'' the two extremes given by a greedy maximum and generous maximum matching. This can be seen in terms of both the percentage of students with first-choice projects and the degree of the matchings. In terms of the percentage of students with first-choice projects, the results show that minimum cost maximum matchings lie almost halfway between greedy and generous maximum matching percentages. In terms of the degree of the matchings, it seems that minimum cost maximum matchings are a lot closer to generous than greedy maximum matchings. This is usually seen as a desirable property.

%--------------------------------------------------------------------------------------------------------------------------------------------------
\section{Conclusion}
\label{ch7_conclusion}
In this paper we investigates the Student / Project Allocation problem in the context of profile-based optimality. We showed how greedy and generous maximum matchings can be found efficiently using network flow techniques. We also presented a range of empirical results obtained from evaluating these efficient algorithms. An obvious question to ask at this stage relates to which other extensions of \scp{spa} of practical relevance or theoretical significance can be investigated. 
These include:
\begin{enumerate}

\item Can we improve on the $O(n_1^2Rm_2)$ algorithm for finding greedy and generous maximum matchings in \scp{spa}? One approach would be to determine whether there  are faster ways of finding maximum profile augmenting paths in the underlying network than that presented in Algorithm \ref{get_auth}. Another approach may be perhaps to abandon the network flow method and consider adopting other techniques used for solving similar problems in the \scp{chat} context \cite{Irv06,MM06,HK13}.

\item The notion of \emph{Pareto optimality} has been well studied in the \scp{ha} context \cite{AS98,ACMM04}. It is easy to see that the profile-based optimality criteria defined here imply Pareto optimality. However studying Pareto optimality in its own right is of theoretical interest. Since Pareto optimal matchings in \scp{chat} can be of varying sizes, this extends to \scp{spa}. Given a \scp{spa} instance we may seek to find a maximum Pareto optimal matching in time faster than  $O(n_1^2Rm_2)$.
\end{enumerate}

%================================================================================================================
%================================================================================================================

%\section*{Bibliography}
%\bibliographystyle{elsarticle-num}
%\bibliographystyle{plain}
%\bibliography{matching_db}

\end{document}